\documentclass[letter]{aa} % for the letters 
\usepackage{graphicx}
\usepackage{txfonts}
\usepackage[colorlinks=true, linkcolor=blue, urlcolor=blue, citecolor=blue, filecolor=blue]{hyperref}
\usepackage{orcidlink}
\usepackage{paralist}
\usepackage[inline]{enumitem}
\setlist{nolistsep}
\setlist[enumerate]{itemsep = 0mm}
\setlist[itemize]{itemsep = 0mm}
\usepackage{amsmath}
\usepackage{nicefrac}
\usepackage{tikz}
\graphicspath{{./}{figures/}}

\begin{document}
%-----------------------------------------------------------------------
\title{First determination of the angular dependence of rise and decay times of solar radio bursts using multi-spacecraft observations}
\subtitle{}

\titlerunning{Decay and Rise Time vs Angle}
\authorrunning{Chrysaphi et al.}

\author{
            Nicolina Chrysaphi \orcidlink{0000-0002-4389-5540} \inst{\ref{inst1},\ref{inst2}, \ref{inst3}}
        \and
            Milan Maksimovic \orcidlink{0000-0001-6172-5062} \inst{\ref{inst1}} 
        \and
            Eduard P. Kontar \orcidlink{0000-0002-8078-0902} \inst{\ref{inst3}} 
        \and
            Antonio Vecchio \orcidlink{0000-0002-2002-1701} \inst{\ref{inst4},\ref{inst1}}
        \and
            Xingyao Chen \orcidlink{0000-0002-1810-6706} \inst{\ref{inst3}} 
        \and
            Aikaterini Pesini \inst{\ref{inst4}} 
          }

\institute{
            LESIA, Observatoire de Paris, Universit\'{e} PSL, CNRS, Sorbonne Universit\'{e}, Universit\'{e} de Paris, 5~place Jules Janssen, 92195 Meudon, France \label{inst1}
            %\email{}
        \and
            Sorbonne Universit\'{e}, \'{E}cole Polytechnique, Institut Polytechnique de Paris, CNRS, Laboratoire de Physique des Plasmas (LPP), 4~Place Jussieu, 75005 Paris, France \label{inst2}
        \and
            School of Physics \& Astronomy, University of Glasgow, Glasgow, G12 8QQ, UK \label{inst3}
        \and
            Radboud Radio Lab, Department of Astrophysics, Radboud University Nijmegen, The Netherlands \label{inst4}
             }

\date{Received 6 October 2023; Accepted 18 February 2024}

%-----------------------------
  \abstract{
  A large arsenal of space-based and ground-based instruments is dedicated to the observation of radio emissions, whether they originate within our solar system or not.  Radio photons interact with anisotropic density fluctuations in the heliosphere, which can alter their trajectory and influence properties deduced from observations.  This is particularly evident in solar radio observations, where anisotropic scattering leads to highly-directional radio emissions.  Consequently, observers at varying locations will measure different properties, including different source sizes, source positions, and intensities.  However, it is not known if measurements of the decay time of solar radio bursts are also affected by the observer’s position.  Decay times are dominated by scattering effects, and so are frequently used as proxies of the level of density fluctuations in the heliosphere, making the identification of any location-related dependence crucial.  We combine multi-vantage observations of interplanetary Type III bursts from four non-collinear, angularly-separated spacecraft with simulations, to investigate the dependence of both the decay- and rise-time measurements on the separation of the observer from the source.  We propose a function to characterise the entire time profile of radio signals, allowing for the simultaneous estimation of the peak flux, decay time, and rise time, while demonstrating that the rise phase of radio bursts has a non-constant, non-exponential growth rate.  We determine that the decay and rise times are independent of the observer's position, identifying them as the only properties to remain unaffected, thus not requiring corrections for the observer's location.  Moreover, we examine the ratio between the rise and decay times, finding that it does not depend on the frequency.  Therefore, we provide the first evidence that the rise phase is also significantly impacted by scattering effects, adding to our understanding of the plasma emission process.
  }
%-----------------------------


\maketitle

%-----------------------------------------------------------------------
\section{Introduction} \label{sec:intro}
Solar radio bursts often result from electron oscillations (Langmuir waves) with frequencies $\gtrsim f_{pe}$ (the local plasma frequency; \cite{2017RvMPP...1....5M}), directly relating solar-radio waves to the local electron density, making them key diagnostics of the local heliospheric environment \citep[][]{2021NatAs...5..796R, 2023ApJ...956..112K}. 
Propagation effects, including scattering from small-scale density fluctuations, distort radio waves -- whether solar or extra-solar in origin \citep{1958MNRAS.118..534H, 2023ApJ...956..112K} -- traversing the heliospheric medium, severely impacting frequencies $f$ close to $f_{pe}$ \citep{1971A&A....10..362S}.
Lower frequencies are affected to a larger extent \citep{2018ApJ...868...79C}, for example, observed solar-emission time profiles (``light curves'') broaden with decreasing frequency, corresponding to longer decay phases (signals after the peak), and thus longer decay times \citep{1969SoPh....8..388A, 1973SoPh...30..175A, 1975SoPh...45..459B, 2018ApJ...857...82K, 2018A&A...614A..69R, 2019ApJ...884..122K}.

The interplay between various radio-wave propagation effects was explored using 3D ray-tracing simulations providing a description of the anisotropy of density fluctuations \citep{2019ApJ...884..122K}.
It was demonstrated that scattering dominates the observed radio-source properties when $f \approx f_{pe}$ \citep{2017NatCo...8.1515K}, and that anisotropic scattering must be considered to simultaneously reproduce multiple observed properties \citep{2019ApJ...884..122K, 2020ApJ...898...94K, 2020ApJ...905...43C, 2023ApJ...956..112K}.
This anisotropy leads to highly-directional emissions, meaning the detector's location may influence observations.
Simulations show that the angular separation $\theta$ between the radio source and detector impacts the observed flux, source size, and position \citep{2019ApJ...884..122K, 2020ApJ...898...94K, 2020ApJ...905...43C, 2021A&A...656A..34M}.  With increasing $\theta$, imaged sizes decrease and sources appear farther from the Sun, whereas the recorded flux can vary by orders of magnitude.  However, it has not been investigated whether decay-time measurements also vary with $\theta$.

Decay times are particularly significant as they are predominantly defined by scattering effects \citep{2019ApJ...873...33B, 2019ApJ...884..122K, 2020ApJ...905...43C}, providing a proxy for the level of density fluctuations $\delta n/n$ that impact the observed properties \citep[][]{2018ApJ...857...82K, 2020ApJS..246...57K}.
Should decay times vary with the detector's position, a significant (angle-dependent) uncertainty has been neglected in previous estimations of $\delta n/n$, potentially altering interpretations.

Decay times tend to be approximated by a single-exponential fit to the recorded signal's decay phase \citep[][]{1972A&A....19..343A, 1973SoPh...31..501E, 1975SoPh...45..459B, 2018ApJ...857...82K, 2020ApJS..246...57K, 2021A&A...656A..33V}.
However, this does not always successfully describe the decay phase's shape, often failing to characterise the signal peak and tail \citep{1972A&A....19..343A, 1975SoPh...45..459B, 2018A&A...614A..69R, 2018ApJ...857...82K, 2020ApJS..246...57K}. Such functions are also sensitive to the temporal resolution and noise of the measurements, and the choice of temporal range considered for the fit.

In this study, we combine multi-spacecraft measurements of Type III solar radio bursts, recording the same event at varying angular separations.  Frequently-observed and covering a large range of frequencies extending from the Sun to $>$1~au, Type III bursts are considered key diagnostic tools of accelerated electrons and density fluctuations in the heliosphere \citep[e.g.][]{2021NatAs...5..796R, 2023ApJ...956..112K}.  We present a function that allows us to fit the entirety of the time profile, improving the estimation and reliability of the decay time, and providing a simultaneous estimation of the rise time and peak flux. We further examine whether the decay and rise times vary with the position of the detector, and investigate the ratio between the two and their relation to the observed frequency.

%-----------------------------------------------------------------------
\section{Multi-vantage observations} \label{sec:obs}
To investigate the angular dependence of rise- and decay-time measurements, we consider radio bursts that were simultaneously observed by multiple detectors with various angular separations between them.
Data was gathered from 4 different space-based radio instruments (detailed in Appendix~\ref{sec:selection_criteria}): RPW onboard Solar Orbiter (SolO; \cite{2020A&A...642A..12M}), FIELDS onboard Parker Solar Probe (PSP; \cite{2016SSRv..204...49B}), SWAVES onboard the Solar TErrestrial RElations Observatory (STEREO-A; hereafter STA; \cite{2008SSRv..136..487B}), and WAVES onboard WIND \citep{1995SSRv...71..231B}.
The recent launches of SolO and PSP are particularly advantageous for this study since they increase the number of solar-dedicated radio instruments orbiting the Sun.  Additionally, WIND is located at L1 (at $\sim$1~au), thus measurements can be considered the same as those of Earth-based instruments, if identical frequencies could be recorded.

Since the instruments' bandwidths and spectral resolution vary, it is not always possible to examine the light curves at the exact same frequency.  Instead, comparable frequencies are selected (Appendix~\ref{sec:selection_criteria}), introducing additional (accounted for) uncertainties in rise- and decay-time measurements due to the frequency-dependent, scattering-induced broadening of time profiles (Sect.~\ref{sec:decay_rise_vs_angle}).

Strict event selection criteria were applied (Appendix~\ref{sec:selection_criteria}) -- leading to nine Type III bursts -- including at least 3 spacecraft observing each burst, and one spacecraft recording the burst's Langmuir waves in situ.  The latter means that the spacecraft is embedded in the radio source, allowing us to approximate the source and spacecraft as co-spatial.

%-----------------------------------------------------------------------
\section{Analysis and results} \label{sec:results}
\begin{figure}[ht!]
    \centering
    \includegraphics[width=0.5\textwidth, keepaspectratio=true]{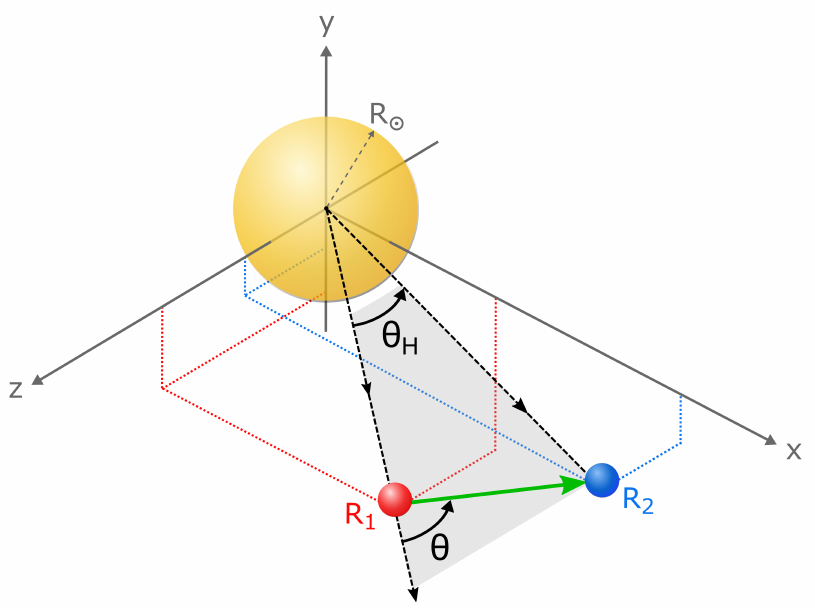}
    \caption{Schematic illustration of the angular separation $\theta$ calculated in the plane of the two spacecraft (grey shaded area), from the Sun-source axis to the vector connecting the spacecraft (green arrow).  The heliocentric angle $\theta_H$ is also shown for comparison.  The red dot at $R_1$ represents the spacecraft observing the Langmuir waves, and thus the radio source location, whereas the blue dot at $R_2$ represents the other spacecraft.  The origin of the 3D Cartesian coordinate system is at the solar centre.
    }
    \label{fig:angle_cartoon}
\end{figure}

The 3D spacecraft positions are used to estimate their angular separation $\theta$ from the radio source (Fig.~\ref{fig:angle_cartoon}; Appendix~\ref{sec:angle_calculation}).
The spacecraft that records Langmuir waves in situ (shown at $R_1$) is considered to be co-spatial with the radio source, defining the Sun-source axis and $\theta=0\degr$.
A second spacecraft at $R_2$ then has $\theta$ measured between this axis and the vector $\vec{R_{12}}$, where $\left|\vec{R_{12}}\right|$ defines the Euclidean distance between the two spacecraft.

%------------------------------------FIGURE-------------------------------------
%----------------------------------END FIGURE-----------------------------------

Notably, the frequently-used heliocentric angle ($\theta_H$ in Fig.~\ref{fig:angle_cartoon}) is not physically meaningful (see Appendix~\ref{sec:angle_calculation}).  Instead, it is angle $\theta$ that needs to be considered, as the angular displacement of a detector from the radio source should be defined according to the maximum of the (scattered) photon beam.

%=================================
\subsection{Fitting the entire light curve} \label{sec:full_curve_fit}
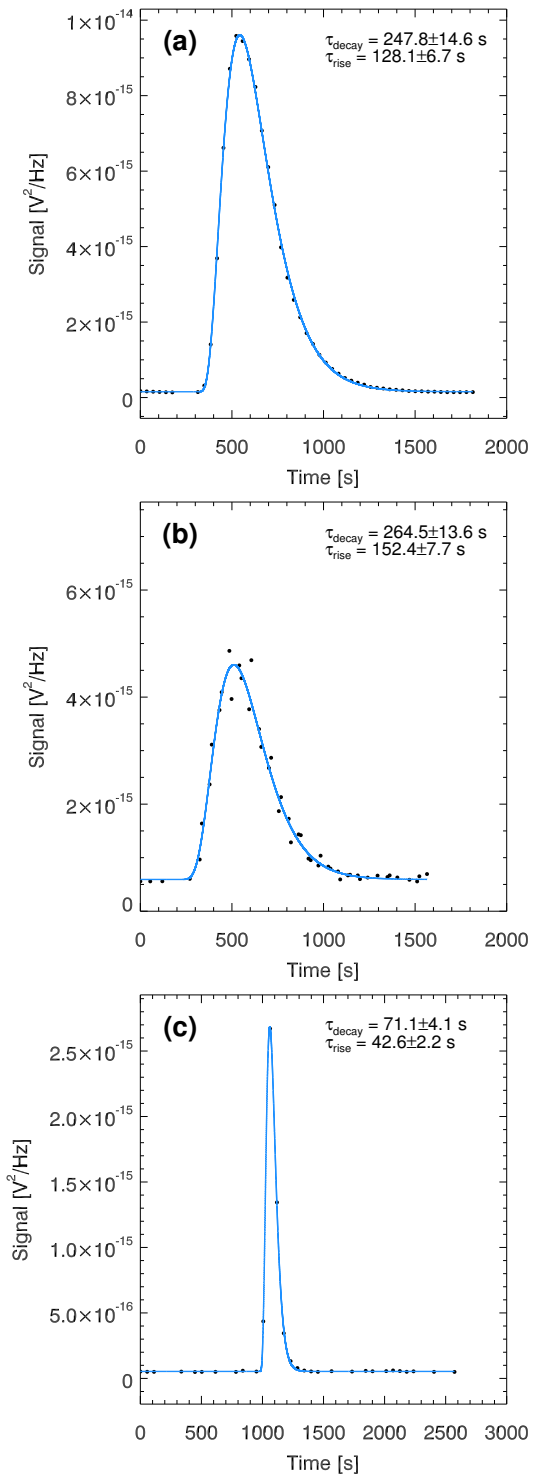
\begin{figure}[ht!]
    \centering
    \begin{tikzpicture}
        \node[anchor=north west,inner sep=0pt] at (-2.3,0.6){
            \includegraphics[height=6.5cm, keepaspectratio=true]{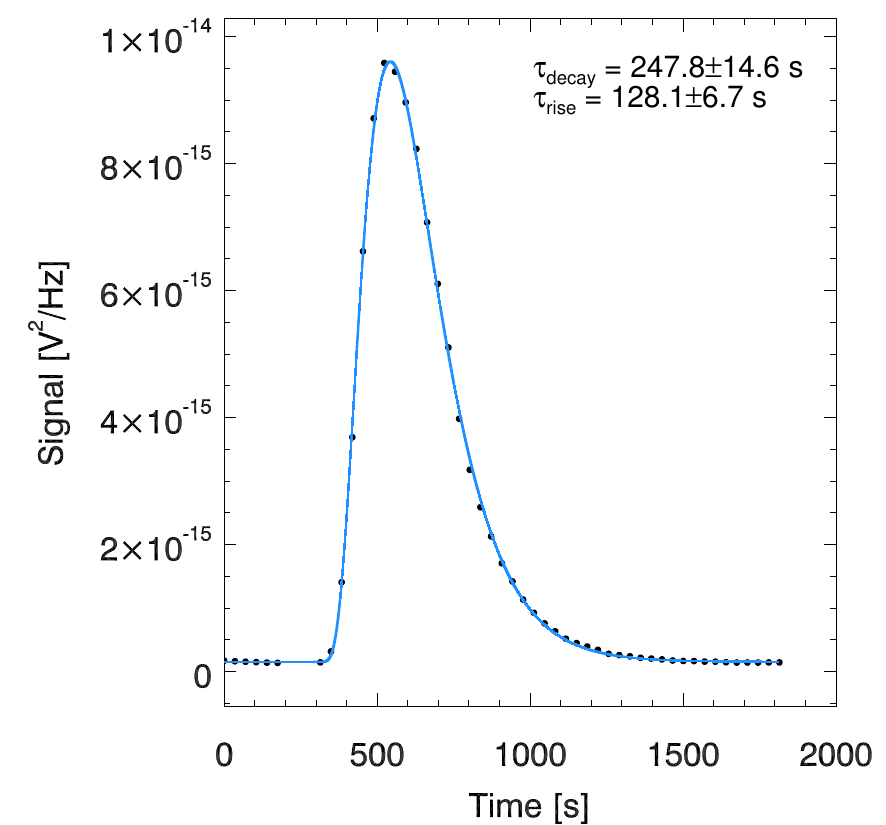}
        }; \node[font=\sffamily\bfseries\large] at (0ex,0ex) {(a)};
    \end{tikzpicture}
    \begin{tikzpicture}
        \node[anchor=north west,inner sep=0pt] at (-2.3,0.6){
            \includegraphics[height=6.5cm, keepaspectratio=true]{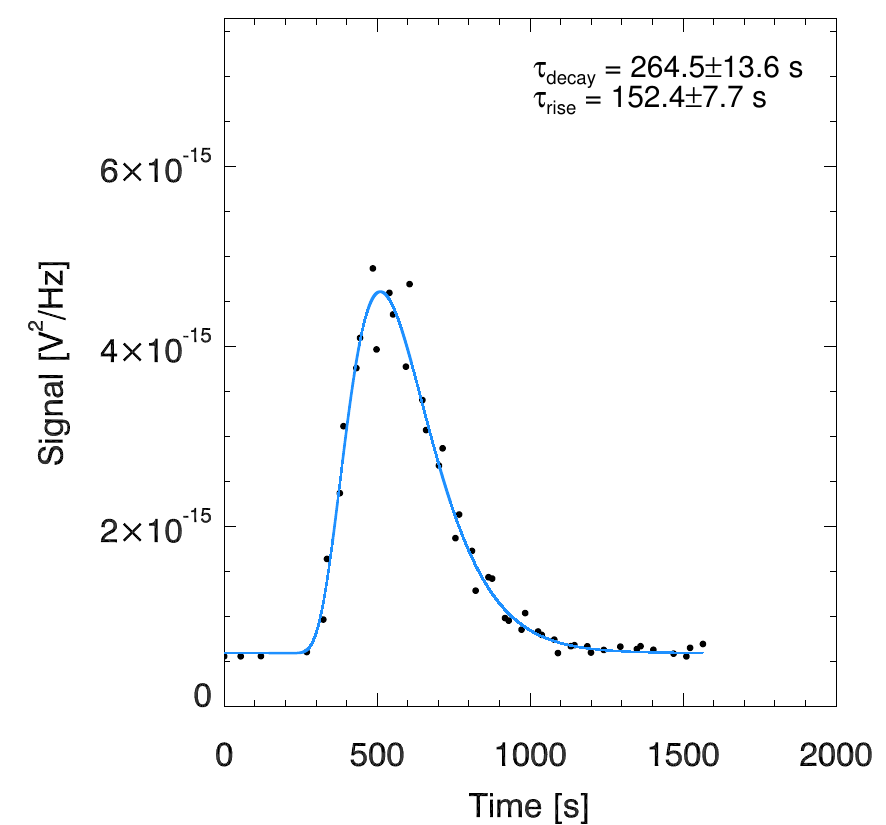}
        }; \node[font=\sffamily\bfseries\large] at (0ex,0ex) {(b)};
    \end{tikzpicture}
    \begin{tikzpicture}
        \node[anchor=north west,inner sep=0pt] at (-2.3,0.6){
            \includegraphics[height=6.5cm, keepaspectratio=true]{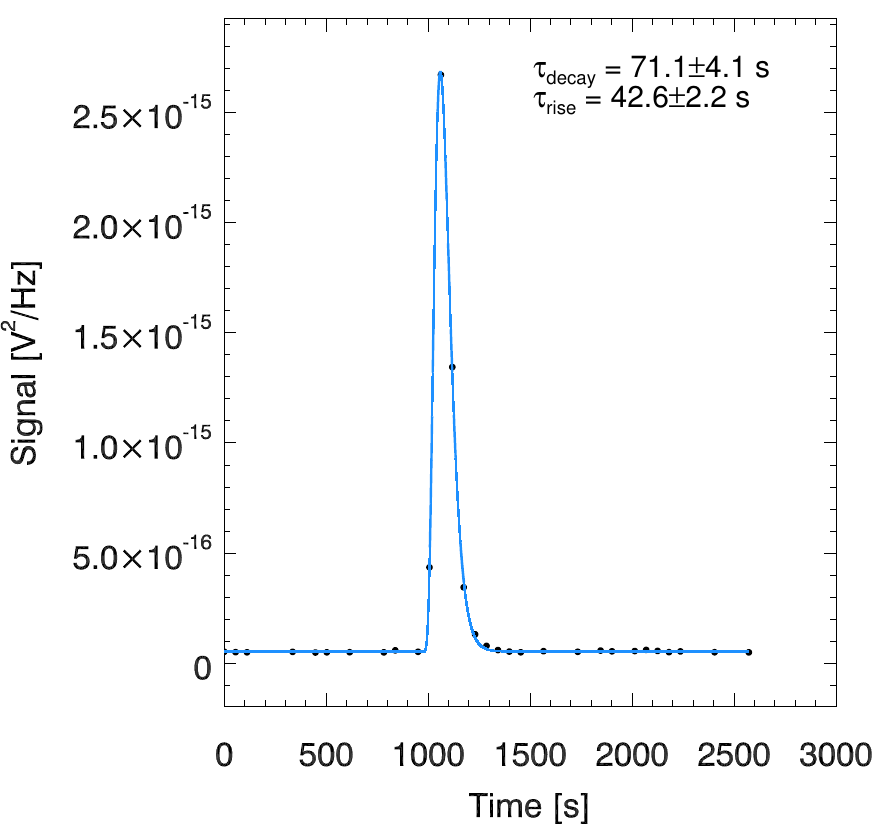}
        }; \node[font=\sffamily\bfseries\large] at (0ex,0ex) {(c)};
    \end{tikzpicture}
    \caption{Single fit of the entire light curve, shown for various frequencies and temporal resolutions.  The blue curve is the fit and the black points are measurements of interplanetary Type III solar radio bursts.  The inferred decay and rise times and their associated errors are indicated in the legend.
    (a) STA data recorded at 325~kHz around 17:50~UT on 18-Nov-2020.
    (b) SolO data recorded at 331.31~kHz around 17:50~UT on 18-Nov-2020.
    (c) PSP data recorded at 1621.88~kHz around 06:00~UT on 22-Aug-2021, included here as an illustration, despite not being used in analyses.
    The presented figures reflect the fitting function's success for 3 different dataset conditions: (a) a smooth and well-resolved signal constituting an ''ideal'' dataset, (b) a faint and noisy signal, and (c) a higher-frequency signal that is not well-resolved.
    }
    \label{fig:fit_function}
\end{figure}

Solar-radio-burst time profiles are asymmetric, smoothly transitioning between the rise and decay phases (Fig.~\ref{fig:fit_function}).
The decay phase tends to be more gradual than the steeper rise phase, a characteristic attributed both to the velocity dispersion of the electron beam and (predominantly) to the scattering that photons experience before reaching the detector \citep{2020ApJ...905...43C}.
As such, decay times are a proxy of radio-wave scattering from heliospheric density fluctuations \citep{2018ApJ...857...82K, 2020ApJS..246...57K}.

The norm for estimating decay times has been to fit decay-phase intensities $I_d$ (i.e. signals measured after the peak-amplitude time $t_0$) with a single exponential function \citep{2021A&A...656A..33V}
\begin{equation} \label{eqn:single_exp}
    I_d = I_{max} \, \exp{\left(- \frac{(t_d-t_0)}{\tau_d}\right)} \, ,
\end{equation}
where $t_d$ is the time, $I_{max}$ is the peak-signal amplitude, and $\tau_d$ is the decay time.
Equation~\ref{eqn:single_exp} requires that $t_d > t_0$, such that -- by definition -- the peak time and amplitude cannot be characterised.
It is indeed not uncommon for a larger time interval after the peak, where the signal is often non-exponential, to be excluded \citep{1972A&A....19..343A, 1975SoPh...45..459B}.  There is, however, no unambiguous motivation for ignoring the initial part of the decay phase, which can impact the estimated decay time (Appendix~\ref{sec:func_reliability}).
Moreover, single-exponential fits often fail to successfully reproduce lower-amplitude signals, which can also influence the decay-time estimations.
Such discrepancies introduce additional (and unaccounted for) errors in decay-time estimations (Appendix~\ref{sec:func_reliability}), and subsequent uncertainties when comparing decay-time measurements, especially between instruments or studies.

Consequently, it is fundamental that functions characterising the entirety of the radio burst time profiles are identified and utilised in analyses.  Here, we propose the following function that successfully reproduces the entirety of radio-burst light curves (Fig.~\ref{fig:fit_function}), allowing for an improved, more reliable (see Appendix~\ref{sec:func_reliability}), and simultaneous estimation of the decay time, the peak-signal amplitudes, and the rise time:
\begin{equation} \label{eqn:fit_function}
    S(t) = \left[ A \, \exp{\left( - \frac{\tau_1}{t-t_{off}} - \frac{t-t_{off}}{\tau_2} \right)} + C \right] \times H(t, t_{off}) \, ,
\end{equation}
where $S(t)$ is the observed signal amplitude (i.e. flux), $t$ represents the observed time, $t_{off}$ is a time offset defining the activation of the Heaviside function $H(t, t_{off})$, $A$ is a pre-factor related to the signal amplitude (such that the observed peak signal $S_{max}=A$ at the peak-amplitude time $t_{peak} = \sqrt{(\tau_1 \, \tau_2)}+t_{off}$), $C$ is a constant offset, and $\tau_1$ and $\tau_2$ are proxies for the rise and decay times, respectively, and must be positive to be physically meaningful.  The condition $(t-t_{off}) > 0$ must be satisfied, which is achieved by the Heaviside function.
The physical rise ($\tau_r$) and decay times ($\tau_d$) can be calculated analytically.  In this study, the $S_{max}/e$ level is used to define both $\tau_r$ and $\tau_d$, such that $\tau_r$ is the time taken to increase from  $S_{max}/e$ to $S_{max}$, and $\tau_d$ is the time taken for the signal to reach $S_{max}/e$ during the decay phase.
Data points below the background-signal level (see Appendix~\ref{sec:bg_estimation}) were excluded from the fit, rather than merely subtracting a constant value.  The magnitude of this background-signal level is accounted for in Eq.~\ref{eqn:fit_function} through $C$.
The inclusion of $H(t, t_{off})$ makes Eq.~(\ref{eqn:fit_function}) discontinuous, therefore, a weighted least-squares fitting procedure with gradient-expansion was applied.
Each free parameter is varied $\pm 10$\% around its optimal value and the corresponding $S_{max}$, $\tau_r$, $\tau_d$, and $t_{peak}$ are obtained for each permutation.  The 1 standard deviation of each property is then defined as the (conservative) uncertainty on the property's value.

It is also noteworthy that the success of Eq.~(\ref{eqn:fit_function}) at characterising the radio time profiles, including their rise phase (Fig.~\ref{fig:fit_function}), demonstrates that the rise phase of radio bursts does not grow exponentially.  A linear exponential function $I_{exp} \propto \exp{(t/\tau)}$ has a constant growth rate ($1/\tau$).
%$\frac{dI_{exp}}{dt} \frac{1}{I_{exp}} = 1/\tau$
To the contrary, the growth rate of the proposed function is not constant, instead, it is a function of time ($dS(t)/dt \propto \tau_1/t^2$), initially growing significantly faster than an exponential, matching the rise phases of radio bursts.

%------------------------------------FIGURE-------------------------------------
%----------------------------------END FIGURE-----------------------------------

%=================================
\subsection{Decay and rise time vs angle} \label{sec:decay_rise_vs_angle}

%------------------------------------FIGURE-------------------------------------
\begin{figure*}[ht!]
    \centering
    \includegraphics[width=0.49\textwidth, keepaspectratio=true]{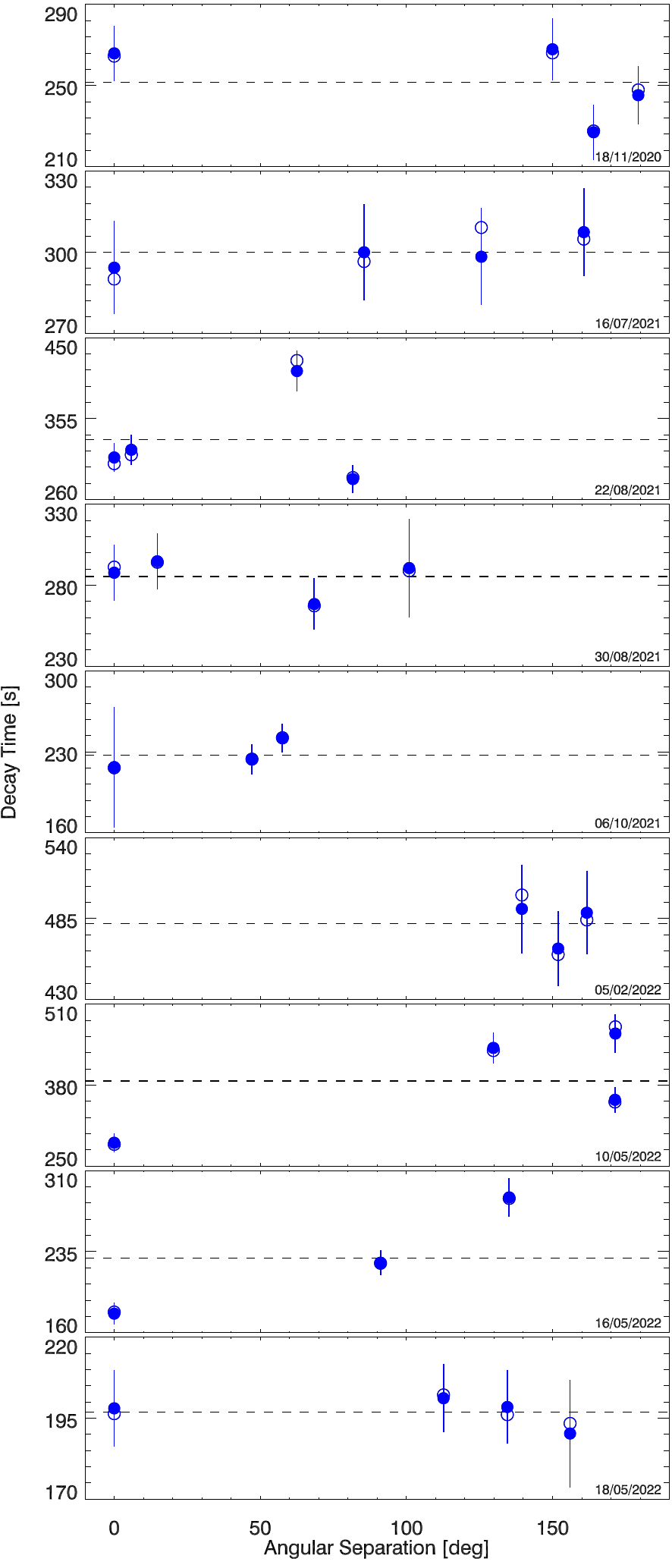}~
    \includegraphics[width=0.49\textwidth, keepaspectratio=true]{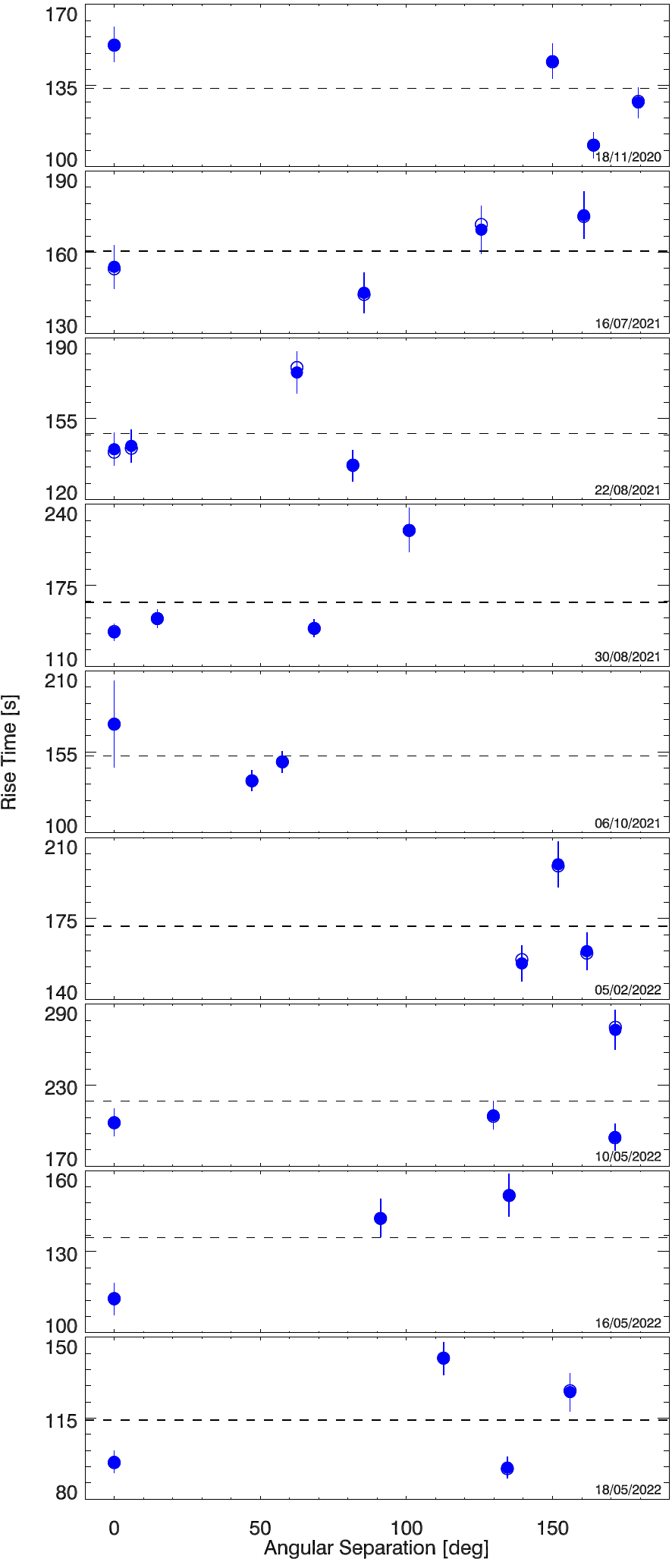}
    \caption{
    Observed decay times (left) and rise times (right), with their associated errors, as a function of the angular separation $\theta$ between the source and the spacecraft.  Each row represents one Type III burst, and the spacecraft that recorded the Langmuir waves in situ is always located at $0\degr$. The black dotted line in each panel represents the average decay/rise time value measured for the given event.  Empty circles represent the decay/rise times obtained by fitting Eq.~(\ref{eqn:fit_function}), whereas filled circles represent the same measurements but corrected for the systematic offset induced by any difference in the recorded frequencies between spacecraft (Appendix~\ref{sec:freq_corrections}).  For clarity, error bars are only plotted for the filled circles.
    }
    \label{fig:decay_rise_vs_angle}
\end{figure*}
%----------------------------------END FIGURE-----------------------------------

%------------------------------------FIGURE-------------------------------------
\begin{figure*}[ht!]
    \centering
    \includegraphics[width=0.47\textwidth, keepaspectratio=true]{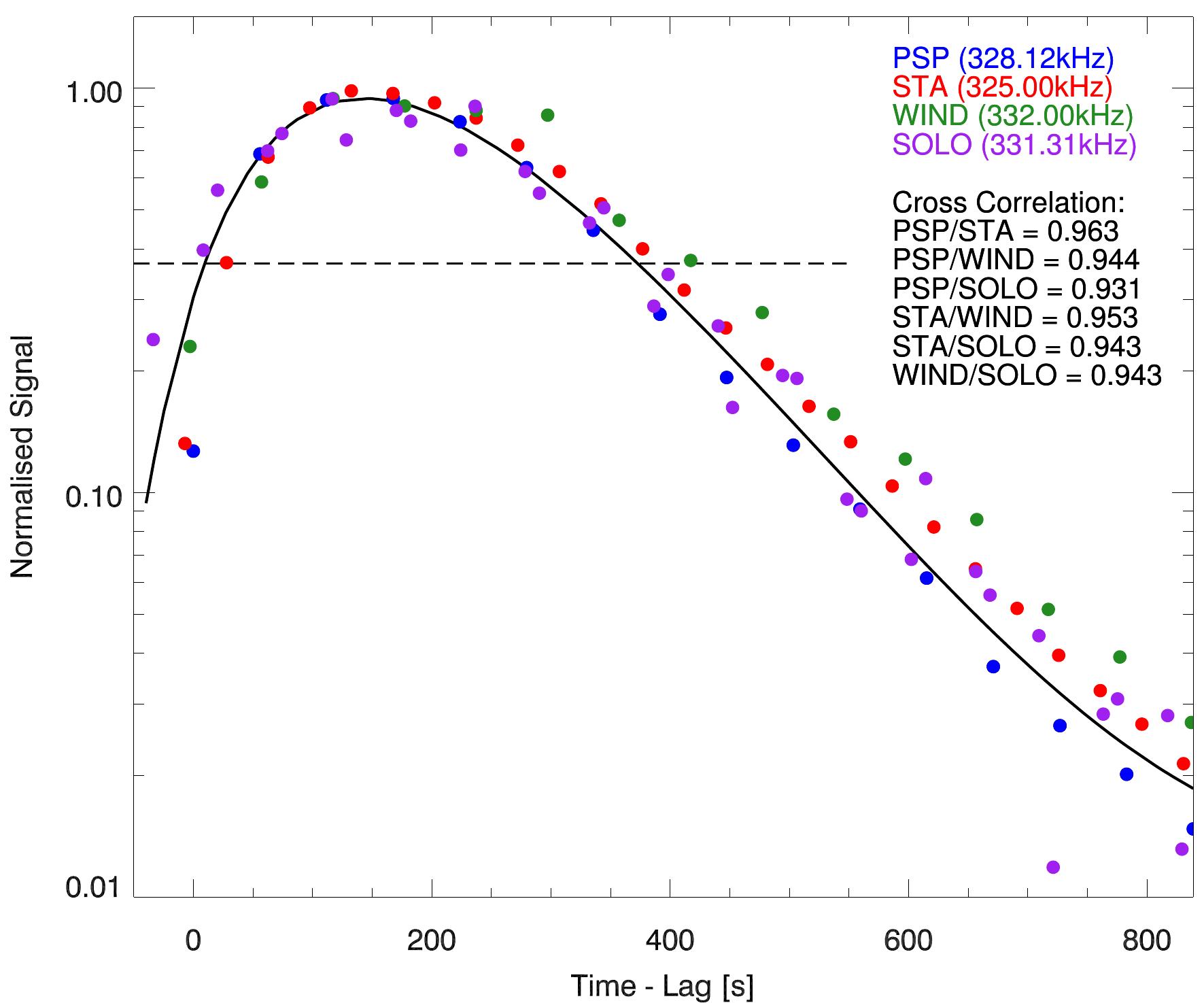}~
    \includegraphics[width=0.47\textwidth, keepaspectratio=true]{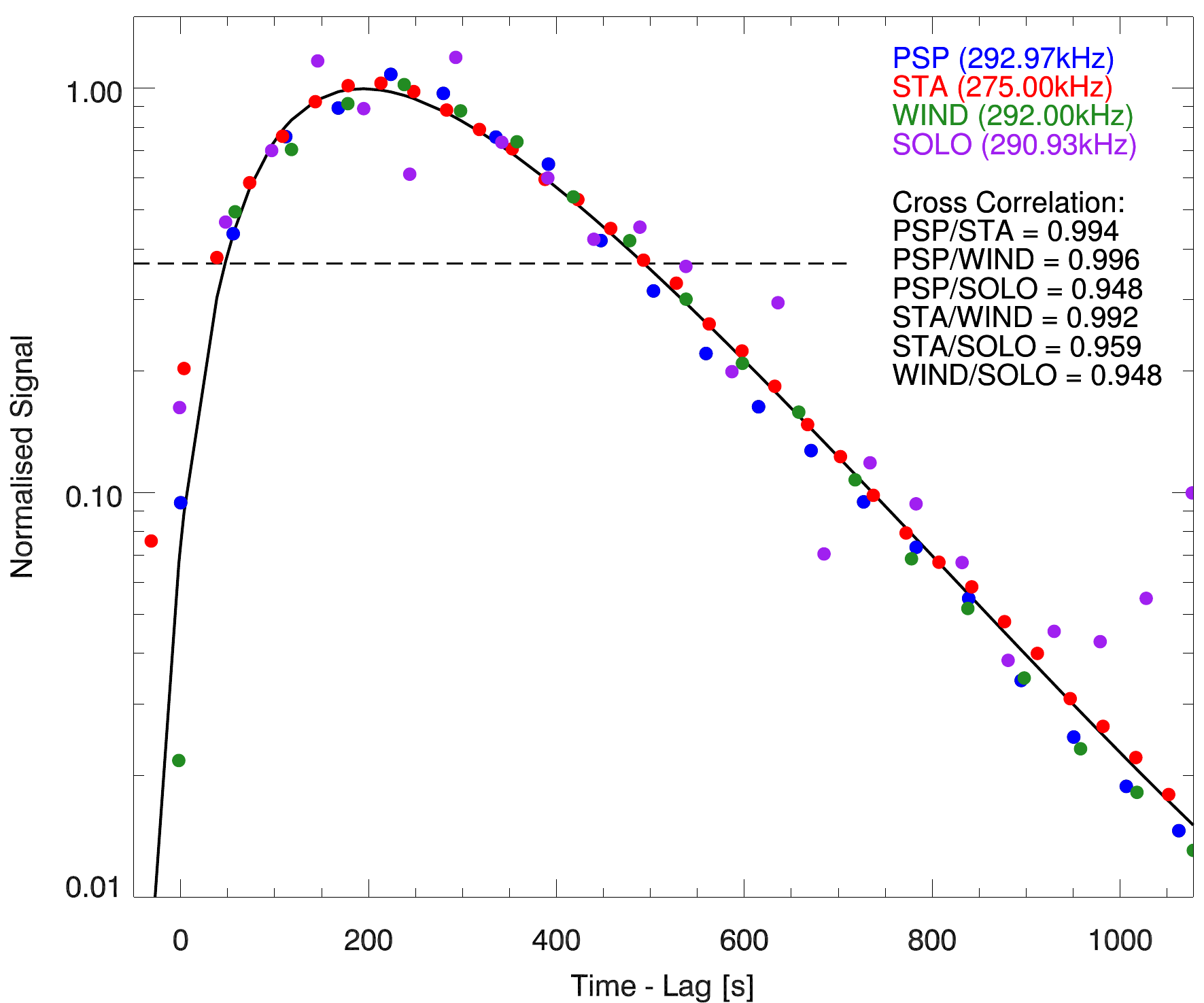}
    \caption{Examples of aligned light curves recorded by four different spacecraft at comparable frequencies, for two separate Type III bursts (shown in each of the panels).
    Light curves were normalised based on their maximum amplitude obtained by fitting, hence the peak data points may appear below or above 1.
    The calculated cross-correlation values between each of the light curves are displayed in the legend of each panel.
    The black dashed line illustrates the $S_{max}/e$ level where the decay and rise times are estimated.
    The black curve is a visual guide, based on applying Eq.~(\ref{eqn:fit_function}) to the data of all the spacecraft in each event simultaneously.
    The depicted events were recorded on 18-Nov-2020 (left) and 16-Jul-2021 (right).
    }
    \label{fig:cross_corr}
\end{figure*}
%----------------------------------END FIGURE-----------------------------------

A strong dependence of radio emission properties (like the observed flux, source position, and size) on $\theta$ has been demonstrated, using both observations and simulations of anisotropic scattering.  However, the angular dependence of decay times -- from which the level of density fluctuations $\delta n/n$ is often inferred -- has not been examined.

First, we invoke 3D ray-tracing simulations which consider anisotropic density fluctuations (see Appendix~\ref{sec:simulations}) in order to obtain a prediction of $\tau_d$ and $\tau_r$ as a function of $\theta$.  These simulations (Fig.~\ref{fig:simulations}) suggest that there is no trend between $\tau_d$ or $\tau_r$ and increasing angular separations.

However, simulations alone do not suffice to draw robust conclusions, so we evaluate their prediction using multi-vantage observations of interplanetary Type III solar radio bursts.
Given the frequency dependence of time profiles, comparable frequencies were analysed in order to eliminate ambiguities between spacecraft observations.
The obtained $\tau_d$ and $\tau_r$ values were corrected for the systematic offset (Appendix~\ref{sec:freq_corrections}) resulting from variations in recorded frequencies between spacecraft observing the same event.

The left (right) panels of Fig.~\ref{fig:decay_rise_vs_angle} depict the decay (rise) times of the observed Type III radio bursts as a function of $\theta$.
The obtained decay and rise times do not show a consistent trend with increasing $\theta$ at comparable frequencies, in agreement with the output of our simulations (Appendix~\ref{sec:simulations}).
Consequently, we identify our first key result: the decay- and rise- times are the only observed solar radio burst properties whose measurements do not consistently vary with the observer's position.

The variation localised to some of the events is likely due to uncertainties in the fitting procedure and the limited (and differing) temporal resolution of the detectors.
A dependency of measured decay times on the observer's distance from the source has been suggested by some studies, like \cite{2021A&A...656A..33V}, however, we find no such dependence (see Appendix~\ref{sec:decay_rise_vs_distance}).

There are two limiting cases where $\tau_d$ may vary between different observers.
First, when measurements at a given frequency are conducted very close to their source, where the majority of scattering takes place.
When the spacecraft is embedded in the radio source, it may not observe the fully-scattered photons, i.e., their distribution may differ compared to distances farther from the source where the initial, dominant scattering has ceased.  This may lead to variations in $\tau_d$ measurements (at comparable frequencies) between spacecraft.  However, frequencies very close to those at which Langmuir waves were recorded were excluded from this analysis (Appendix~\ref{sec:selection_criteria}), to ensure that the measurements of spacecraft closer to the source can be confidently compared.
A second rare case causing $\tau_d$ to vary as a function of $\theta$ has been identified in our simulations, where radio sources are confined in relatively short, highly-curved magnetic field lines close to the surface of the Sun (i.e. loops in a dipole), as is the case for Spike bursts \citep{2023ApJ...946...33C}.  In this extreme case where the photon beam's propagation is strongly constrained by the magnetic field configuration, we find that $\tau_d$ may differ between observers.
However, this is not applicable to Type III bursts (or other interplanetary radio emissions) which are associated with open magnetic field lines, and thus is not relevant here.

Our results are evaluated by cross-correlating the time profiles of each Type III burst recorded by the different spacecraft (Appendix~\ref{sec:cross_correl}).  Figure~\ref{fig:cross_corr} depicts the aligned, normalised light curves of two of the analysed events.  Even a direct comparison of light curves at non-identical frequencies demonstrates that spacecraft at varying angular separations and distances from the source record light curves whose width (and thus $\tau_d$ and $\tau_r$) does not vary significantly.
Therefore, this comparison and the high cross-correlation values obtained (often $\gtrsim$~95\%), corroborate the outcome of the independence of rise- and decay-time measurements on the observer's (angular and Euclidean) separation from the source, as obtained using the proposed fitting function.

%=================================
\subsection{Rise-to-decay time ratio} \label{sec:rise_decay_ratio}
\begin{figure}[ht!]
    \centering
    \begin{tikzpicture}
        \node[anchor=south east,inner sep=0pt] at (1.1,-1.4){
            \includegraphics[width=0.49\textwidth, keepaspectratio=true]{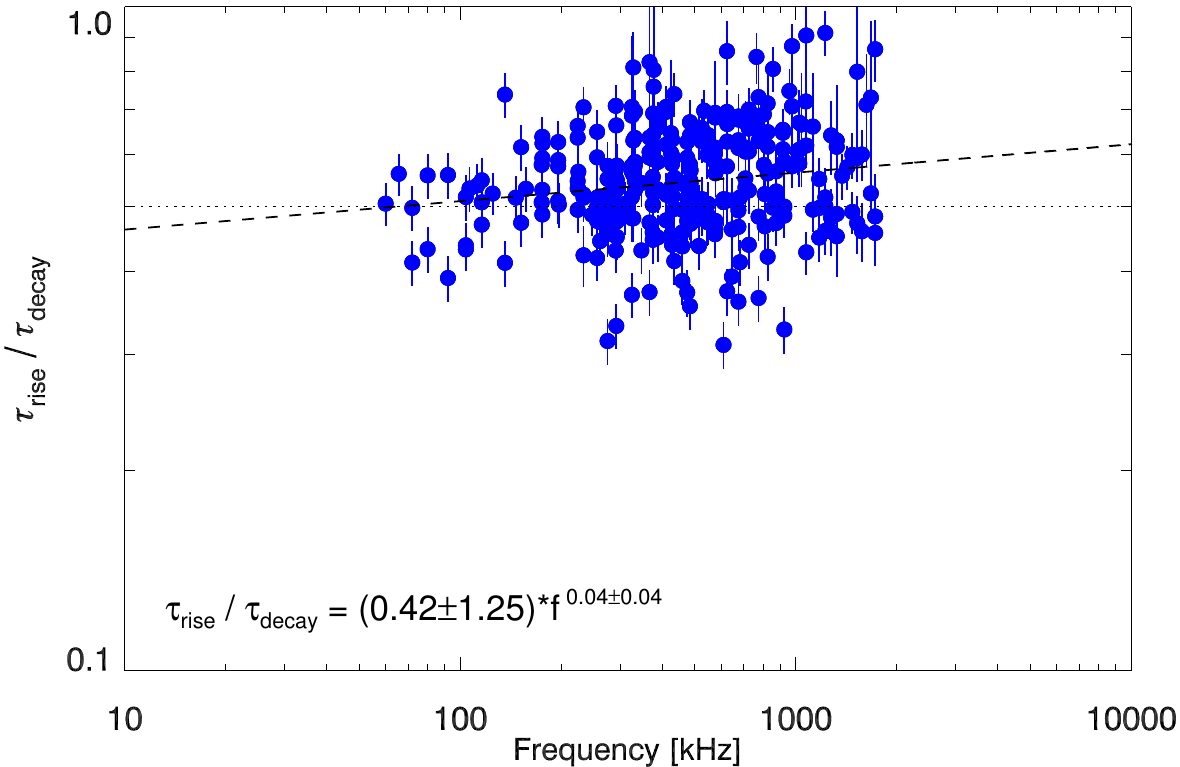}
        }; \node[font=\sffamily\bfseries\large] at (0ex,0ex) {(a)};
    \end{tikzpicture}
    \begin{tikzpicture}
        \node[anchor=north east,inner sep=0pt] at (1.1,0.7){
            \includegraphics[width=0.49\textwidth, keepaspectratio=true]{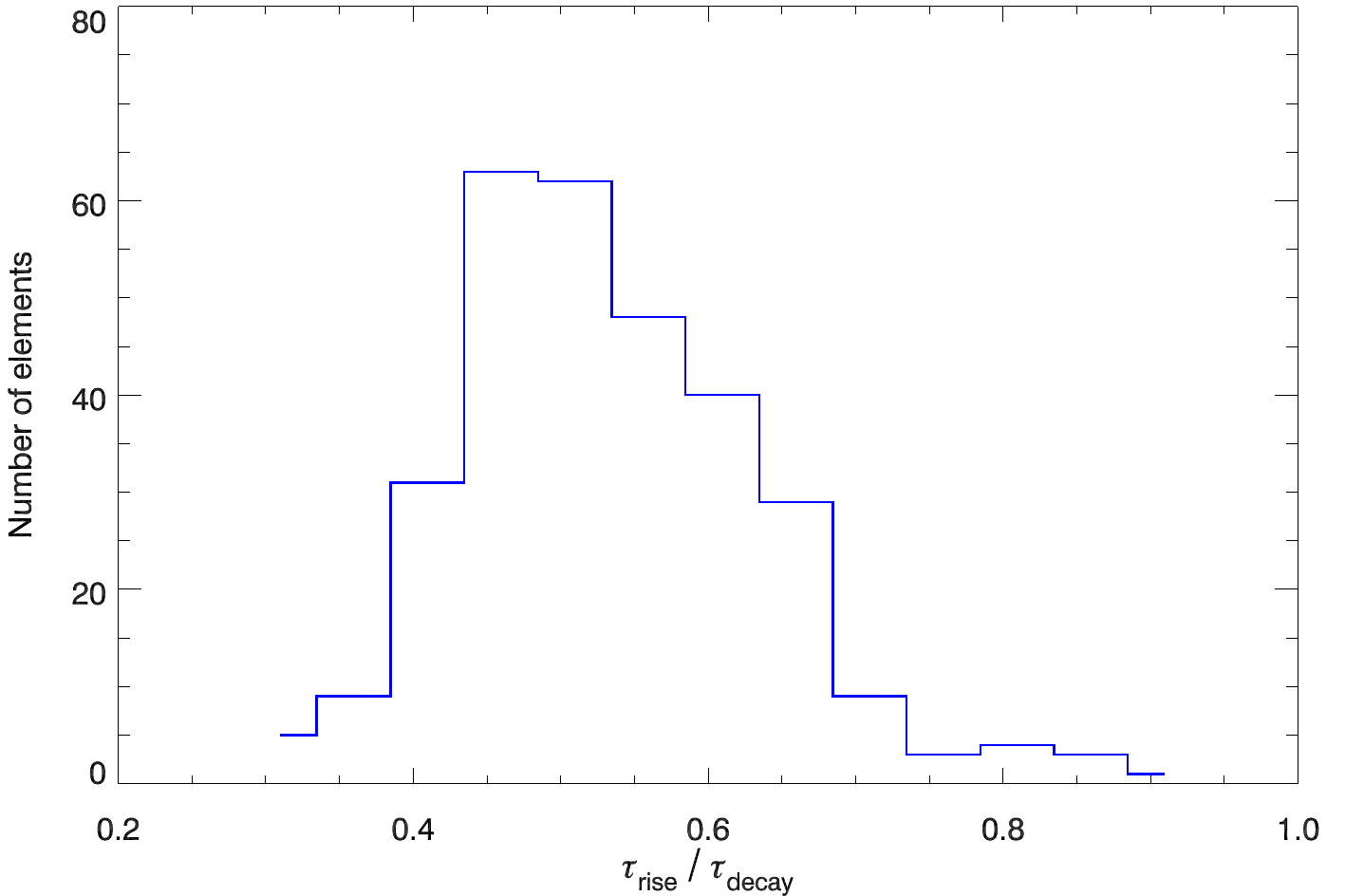}
        }; \node[font=\sffamily\bfseries\large] at (0ex,0ex) {(b)};
    \end{tikzpicture}
    \begin{tikzpicture}
        \node[anchor=north east,inner sep=0pt] at (1.1,0.7){
            \includegraphics[width=0.49\textwidth, keepaspectratio=true]{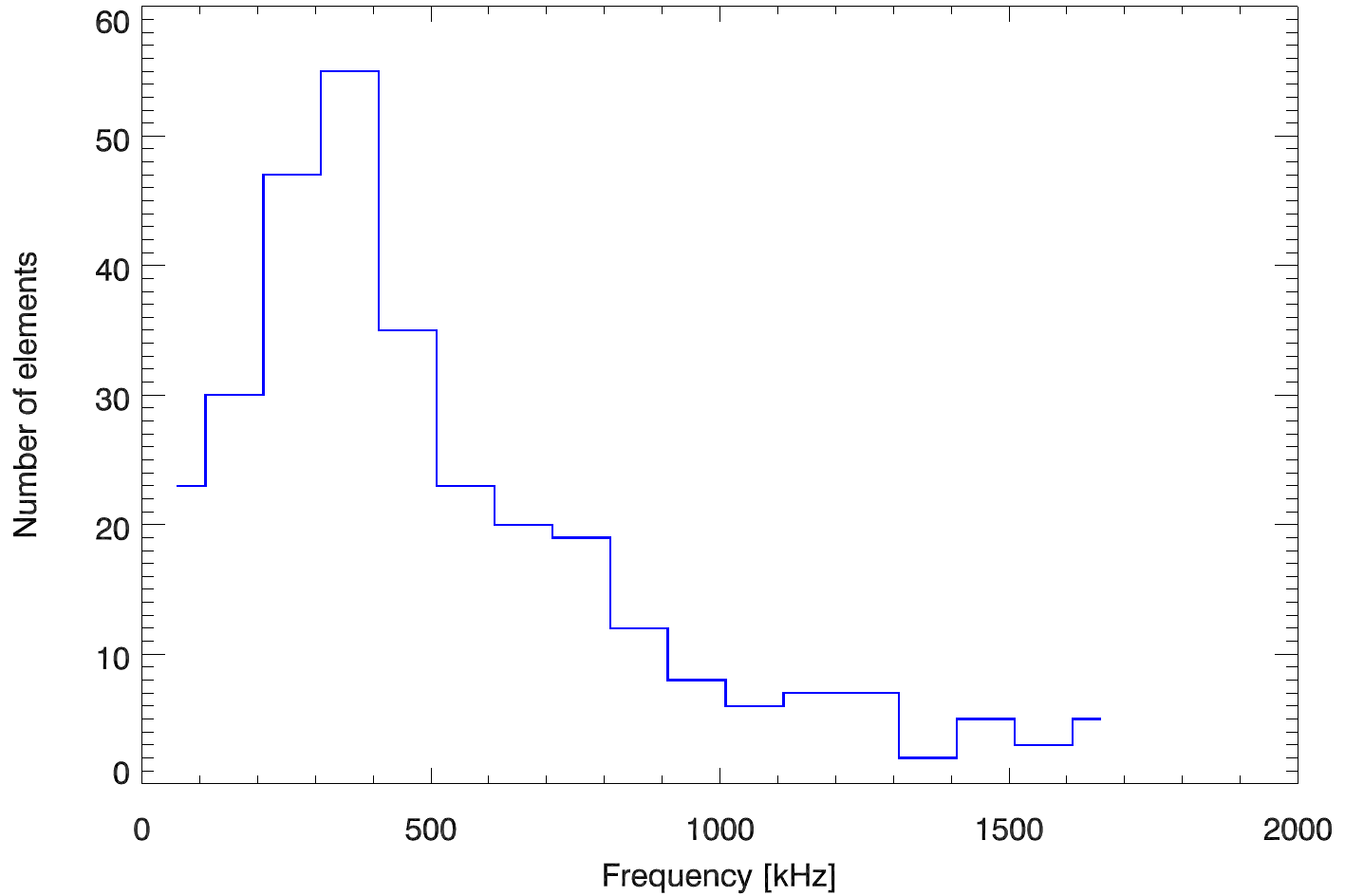}
        }; \node[font=\sffamily\bfseries\large] at (0ex,0ex) {(c)};
    \end{tikzpicture}
    \caption{
    (a) Ratios (and associated uncertainties) of rise- to decay-times with respect to frequency, for 307 measurements at frequencies ranging from 60--1725~kHz.
    The rise and decay times were calculated at the $S_{max}/e$ level using Eq.~(\ref{eqn:fit_function}).
    The black dashed line represents the weighted linear fit to the data, with the obtained relationship stated on the bottom of the panel.  The black dotted line indicates the level where the ratio is equal to 0.5.
    (b) Histogram of the calculated rise-to-decay time ratios, with the median value being $\sim$0.54 (and the average value being $\sim$0.56).
    (c) Histogram of the frequencies analysed for the rise-to-decay time ratio calculations.
    }
    \label{fig:rise_decay_ratio}
\end{figure}
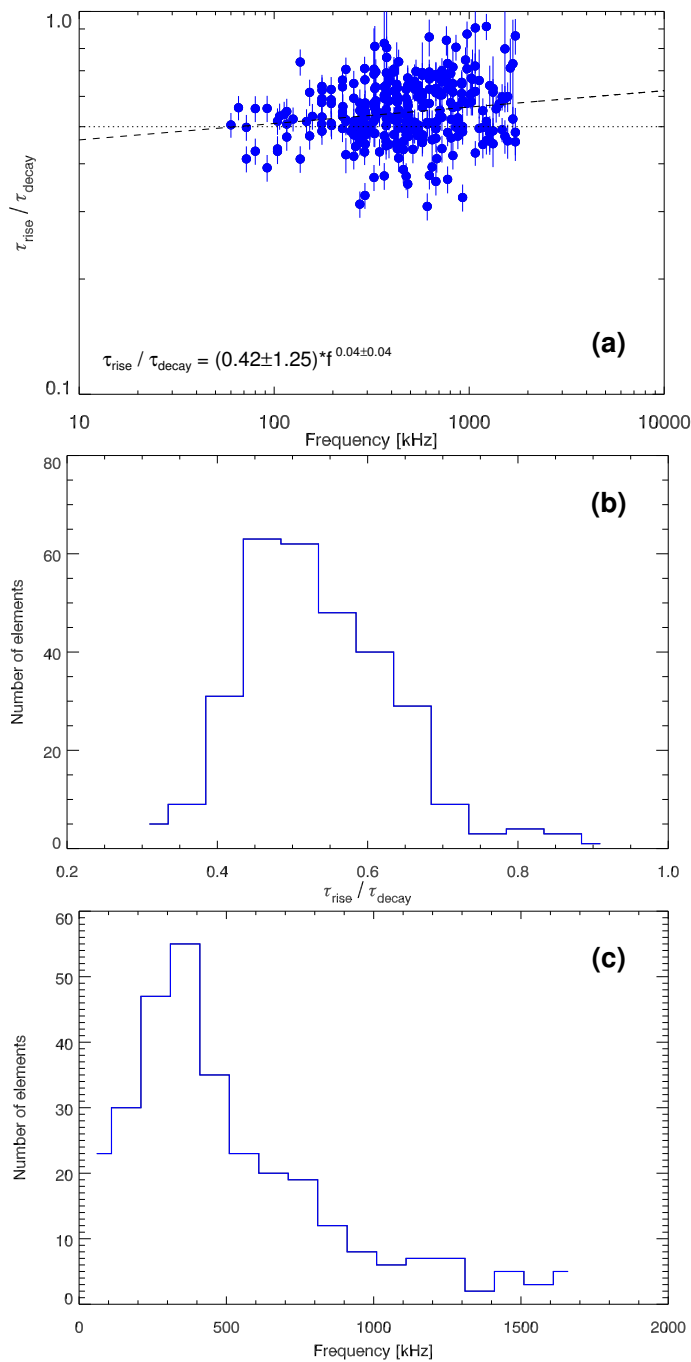

Equation~(\ref{eqn:fit_function}) allows for a simultaneous estimation of $\tau_r$ and $\tau_d$, an important consequence for robust comparisons.
The dependence of both $\tau_r$ and $\tau_d$ as a function of frequency is examined in Appendix~\ref{sec:decay_rise_vs_freq}.
The exact ratio $\tau_r / \tau_d$, however, has been an open question, even though the asymmetry of radio time profiles is well-known.  The obtained $\tau_r / \tau_d$ ratios are presented in Fig.~\ref{fig:rise_decay_ratio}a, as a function of frequency.  It can be seen that, as expected, the rise time is shorter than the decay time, but there is a lack of dependence on the observed frequency, with the obtained relationship being
\begin{equation} \label{eqn:rd_ratio_vs_freq}
    \frac{\tau_r}{\tau_d} \propto f^{ \, 0.04 \pm 0.04} \, .
\end{equation}
The calculated $\tau_r / \tau_d$ values range between $\sim$0.31--0.91, with the majority being between $\sim$0.45--0.65 (Fig.~\ref{fig:rise_decay_ratio}b).
This analysis was conducted using 307 measurements at the $S_{max}/e$ level for frequencies from 60--1725~kHz (Fig.~\ref{fig:rise_decay_ratio}c), obtained from the Type III bursts listed in Table~\ref{tab:events_list}.

%------------------------------------FIGURE-------------------------------------
%----------------------------------END FIGURE-----------------------------------

Our findings can be compared to previous studies, increasing the considered dataset and frequencies.
Meaningful comparisons must ensure that equivalent definitions of the rise and decay times are used, both in terms of the signal level they are measured at, and crucially the time ranges used to define each phase (Appendix~\ref{sec:func_reliability}).
The ratios estimated using Eq.~(\ref{eqn:fit_function}) were re-calculated at the $S_{max}/2$ level (instead of $S_{max}/e$) to enable a comparison with past studies, obtaining values between $\sim$0.37--0.93.
\cite{1972A&A....19..343A} state -- without error analysis -- that the time-profiles' shape varies very little for frequencies between 29.3--130~MHz, with the ratio between $\tau_d$ and the duration ($= \tau_r + \tau_d$) of Type III bursts (measured at $S_{max}/2$) remaining relatively constant (0.55--0.62), and from which we calculate $\tau_r/\tau_d$ to vary from $\sim$0.61 to $\sim$0.82.
\cite{2018A&A...614A..69R} studied the ratio between $\tau_r$ and $\tau_d$ (measured at $S_{max}/2$) for frequencies $\sim$30--70~MHz, finding that $\tau_r/\tau_d$ ranges between $\sim$0.63--0.83.
They suggest (also without error analysis) that the ratio $\tau_r/\tau_d$ increases as the frequency increases, even though a dependence of the estimated ratios on the frequency can only be claimed between $\sim$30--40~MHz, and no dependence appears to exist between $\sim$40--70~MHz \citep[see Fig.~3 of][]{2018A&A...614A..69R}.
Notably, the $\tau_r / \tau_d$ obtained here suggests no clear frequency dependence (given the scatter in the data and the error on the exponent; Eq.~(\ref{eqn:rd_ratio_vs_freq}) and Fig.~\ref{fig:rise_decay_ratio}a), hinting that the time-profile asymmetry is frequency independent.
Moreover, our analysis indicates that even for much lower frequencies, between $\sim$0.06--1.73~MHz, the $\tau_r/\tau_d$ ratios can be as large as $\sim$0.9, contrasting the suggestion of a frequency-dependent ratio.
\cite{2023ApJ...955L..20J} analysed Type III bursts (at the $S_{max}/2$ level) between $\sim$0.75--19~MHz, and even though they give neither $\tau_r/\tau_d$ nor $\tau_d/\tau_r$ as a function of frequency, we can use the information they provide to obtain $\tau_r/\tau_d$  as a function of frequency for the bursts they analysed, enabling another comparison with our results.  Specifically, they found that fundamental bursts have a $\tau_r \propto f^{-0.62 \pm 0.20}$ and $\tau_d \propto f^{-0.73 \pm 0.15}$.  Similarly, they found that harmonic bursts have a $\tau_r \propto f^{-0.48 \pm 0.08}$ and $\tau_d \propto f^{-0.5 \pm 0.06}$.  We note that \cite{2023ApJ...955L..20J} did not consider the fitting errors in estimations of the decay and rise times, and instead give errors that represent the standard deviation of the scatter in their dataset.  Nevertheless, we can calculate that for fundamental emissions $\tau_r/\tau_d \propto f^{\, 0.11 \pm 0.25}$ and for harmonic emissions $\tau_r/\tau_d \propto f^{\, 0.02 \pm 0.1}$.  As such, we find an agreement with our conclusion on the frequency independence of $\tau_r/\tau_d$ (Eq.~(\ref{eqn:rd_ratio_vs_freq}) and Fig.~\ref{fig:rise_decay_ratio}a), irrespective of whether the emissions are fundamental or harmonic, or a convolution of both (Appendix~\ref{sec:selection_criteria}).

The relatively large $\tau_r/\tau_d$ ranges observed (even at comparable frequencies) could reflect differences in the intrinsic radio-emission process.
The exact ratio value may be determined by the size of the emitting region and the intrinsic excitation (and thus injection) time of photons into the heliosphere, where a longer injection from a larger volume causes longer rise times (thus increasing $\tau_r / \tau_d$) for some radio sources. 
The interaction of the electron beam (which is dispersed in space and velocity) and resulting Langmuir waves with density fluctuations populating the heliosphere can impact both the photon-emission region and photon-injection time, contributing to the observed rise phase \citep[e.g.][]{2012SoPh..279..173L, 2014A&A...572A.111R, 2018ApJ...867..158R, 2021NatAs...5..796R}, and thus leading to variations in the $\tau_r/\tau_d$ values between bursts at equivalent frequencies.
Variations in the emission's duration and intrinsic volume may also occur between fundamental ($f \gtrsim f_{pe}$) and harmonic ($f \approx 2 f_{pe}$) sources \citep{2020ApJ...905...43C}.

It is accepted that the observed decay time of solar radio bursts is predominantly defined by the frequency-dependent scattering from small-scale density fluctuations \citep{2018ApJ...857...82K, 2019ApJ...873...33B, 2019ApJ...884..122K, 2020ApJ...905...43C}.
However, the exact contributions to the recorded rise time -- which include the volume of the emitting region and the photon-injection time into the heliosphere -- have not been well understood.
These processes, however, appear insufficient in explaining the observed frequency independence of $\tau_r/\tau_d$, given that they cannot explain the consistent balancing of the scattering-induced broadening of the decay phase with decreasing frequency.
Both the emitting volume and injection time of photons can reasonably be expected to vary between different radio bursts.  Nevertheless, the rise- and decay-times change proportionally with frequency, such that their ratio remains constant.
The independence of $\tau_r/\tau_d$ on the observed frequency hints at the possibility that the rise time is also dictated by scattering.  Therefore, we propose that scattering is a non-negligible contributor to the observed rise times of radio bursts, affecting them in a frequency-dependent and proportionate manner to the decay times.
This should perhaps not come as a surprise, given that scattering affects photons immediately after their injection into the heliosphere, including the photons that form the rise phase (and not selectively only those that form the decay phase).  A hint of the scattering contribution to the rise phase could also be obtained from simulated light curves.  Our simulations account solely for propagation effects, assuming a point source and (crucially) an instantaneous injection of photons \citep{2019ApJ...884..122K}.  Despite that, the resulting light curves are broad, and have a similar shape (both for the rise and decay phases) to those observed \citep[e.g.][]{2020ApJ...905...43C}, alluding to the significant contribution of scattering effects to the observed rise phase.
In other words, in the absence of the contributions from an extended radio source and extended photon-emission duration, radio-wave propagation effects alone can account for a significant part of the rise phase of radio signals.

%-----------------------------------------------------------------------
\section{Conclusions} \label{sec:conclusion}
Using multiple interplanetary Type III solar radio bursts, the dependence of rise- and decay-time measurements on the observer's angular separation from the source was examined.
Each burst was observed by at least 3 non-collinear, angularly-separated space-based instruments onboard Solar Orbiter, Parker Solar Probe, STEREO-A, and WIND.
We propose a function that successfully describes the entirety of the time profiles, allowing for an improved and more consistent calculation of the decay time, and a simultaneous calculation of the rise time and peak-signal amplitude.
The ability to characterise the entire light curve provides improved estimations of the various properties when the data measurements are noisy or when the temporal resolution is low, which can prohibit a successful single-exponential fit and lead to ambiguous results.
Notably, we address the misconception that the rise phase of radio bursts grows exponentially, demonstrating that this is not the case, finding that the rise phase grows non-exponentially at a non-constant rate, where it is initially significantly faster than an exponential growth.

The directivity of solar radio emissions leads to differences in the observed flux, the source size, and the source position, when recorded at different vantage points.
In this study, it was found that neither the decay-time or rise-time measurements at comparable frequencies depend on the angular position of different observers, corroborated in three ways: using simulations, by fitting radio burst observations, and by cross-correlating observations of the same event by different spacecraft.
Furthermore, the decay- and rise-time measurements between spacecraft do not show a systematic dependence on the Euclidean distance from the radio source.
Therefore, the decay and rise times are the only observed properties of radio sources that do not vary with the observer's position, at comparable frequencies.
This result implies that studies using decay times as proxies for estimating the level of density fluctuations $\delta n/n$ in the corona do not need to correct for the detector's position, unlike for other source properties.
We also find no frequency dependence of the rise-to-decay time ratio (based on data between 0.06--130~MHz, i.e. 4 decades of frequency), providing the first evidence that scattering effects impact the duration of the rise phase, as they do for the decay phase of solar radio bursts.
Consequently, we identify scattering effects as an important contributor to the rise time, adding to our understanding of the plasma emission process.

In summary, the work presented here highlights that rise- and decay-time measurements of solar radio bursts can be trusted irrespective of the observer's angular or Euclidean separation from the source.  These two properties can therefore be used as constrains in attempts to infer the true nature of radio sources.

%-----------------------------------------------------------------------
\begin{acknowledgements}
N.C. acknowledges funding support from CNES and from the Initiative Physique des Infinis (IPI), a research training program of the Idex SUPER at Sorbonne Universit\'{e}.
E.P.K. and X.C. were supported by the STFC consolidated grant ST/T000422/1.
The authors would like to thank Dr A. Ross for valuable discussions.
The authors would also like to thank the SolO/RPW, PSP/FIELDS, STEREO/SWAVES, and WIND/WAVES teams for the data.
Solar Orbiter is a mission of international cooperation between ESA and NASA, operated by ESA.
The RPW instrument has been designed and funded by CNES, CNRS, the Paris Observatory, The Swedish National Space Agency, ESA-PRODEX and all the participating institutes.
The FIELDS experiment on the Parker Solar Probe spacecraft was designed and developed under NASA contract NNN06AA01C.
The WIND/WAVES instrument is a joint effort of the Paris Observatory, NASA/GSFC, and the University of Minnesota.
\end{acknowledgements}

%-------------------------------------------------------------------
   \bibliographystyle{aa} % style aa.bst
   \bibliography{references} % your references Yourfile.bib

\begin{thebibliography}{39}
\expandafter\ifx\csname natexlab\endcsname\relax\def\natexlab#1{#1}\fi

\bibitem[{{Alexander} {et~al.}(1969){Alexander}, {Malitson}, \&
  {Stone}}]{1969SoPh....8..388A}
{Alexander}, J.~K., {Malitson}, H.~H., \& {Stone}, R.~G. 1969, \solphys, 8, 388

\bibitem[{{Alvarez} \& {Haddock}(1973)}]{1973SoPh...30..175A}
{Alvarez}, H. \& {Haddock}, F.~T. 1973, \solphys, 30, 175

\bibitem[{{Aubier} \& {Boischot}(1972)}]{1972A&A....19..343A}
{Aubier}, M. \& {Boischot}, A. 1972, \aap, 19, 343

\bibitem[{{Bale} {et~al.}(2016){Bale}, {Goetz}, {Harvey}, {Turin}, {Bonnell},
  {Dudok de Wit}, {Ergun}, {MacDowall}, {Pulupa}, {Andre}, {Bolton},
  {Bougeret}, {Bowen}, {Burgess}, {Cattell}, {Chandran}, {Chaston}, {Chen},
  {Choi}, {Connerney}, {Cranmer}, {Diaz-Aguado}, {Donakowski}, {Drake},
  {Farrell}, {Fergeau}, {Fermin}, {Fischer}, {Fox}, {Glaser}, {Goldstein},
  {Gordon}, {Hanson}, {Harris}, {Hayes}, {Hinze}, {Hollweg}, {Horbury},
  {Howard}, {Hoxie}, {Jannet}, {Karlsson}, {Kasper}, {Kellogg}, {Kien},
  {Klimchuk}, {Krasnoselskikh}, {Krucker}, {Lynch}, {Maksimovic}, {Malaspina},
  {Marker}, {Martin}, {Martinez-Oliveros}, {McCauley}, {McComas}, {McDonald},
  {Meyer-Vernet}, {Moncuquet}, {Monson}, {Mozer}, {Murphy}, {Odom},
  {Oliverson}, {Olson}, {Parker}, {Pankow}, {Phan}, {Quataert}, {Quinn},
  {Ruplin}, {Salem}, {Seitz}, {Sheppard}, {Siy}, {Stevens}, {Summers}, {Szabo},
  {Timofeeva}, {Vaivads}, {Velli}, {Yehle}, {Werthimer}, \&
  {Wygant}}]{2016SSRv..204...49B}
{Bale}, S.~D., {Goetz}, K., {Harvey}, P.~R., {et~al.} 2016, \ssr, 204, 49

\bibitem[{{Barrow} \& {Achong}(1975)}]{1975SoPh...45..459B}
{Barrow}, C.~H. \& {Achong}, A. 1975, \solphys, 45, 459

\bibitem[{{Bian} {et~al.}(2019){Bian}, {Emslie}, \&
  {Kontar}}]{2019ApJ...873...33B}
{Bian}, N.~H., {Emslie}, A.~G., \& {Kontar}, E.~P. 2019, \apj, 873, 33

\bibitem[{{Bonnin} {et~al.}(2008){Bonnin}, {Hoang}, \&
  {Maksimovic}}]{2008A&A...489..419B}
{Bonnin}, X., {Hoang}, S., \& {Maksimovic}, M. 2008, \aap, 489, 419

\bibitem[{{Bougeret} {et~al.}(2008){Bougeret}, {Goetz}, {Kaiser}, {Bale},
  {Kellogg}, {Maksimovic}, {Monge}, {Monson}, {Astier}, {Davy}, {Dekkali},
  {Hinze}, {Manning}, {Aguilar-Rodriguez}, {Bonnin}, {Briand}, {Cairns},
  {Cattell}, {Cecconi}, {Eastwood}, {Ergun}, {Fainberg}, {Hoang}, {Huttunen},
  {Krucker}, {Lecacheux}, {MacDowall}, {Macher}, {Mangeney}, {Meetre},
  {Moussas}, {Nguyen}, {Oswald}, {Pulupa}, {Reiner}, {Robinson}, {Rucker},
  {Salem}, {Santolik}, {Silvis}, {Ullrich}, {Zarka}, \&
  {Zouganelis}}]{2008SSRv..136..487B}
{Bougeret}, J.~L., {Goetz}, K., {Kaiser}, M.~L., {et~al.} 2008, \ssr, 136, 487

\bibitem[{{Bougeret} {et~al.}(1995){Bougeret}, {Kaiser}, {Kellogg}, {Manning},
  {Goetz}, {Monson}, {Monge}, {Friel}, {Meetre}, {Perche}, {Sitruk}, \&
  {Hoang}}]{1995SSRv...71..231B}
{Bougeret}, J.~L., {Kaiser}, M.~L., {Kellogg}, P.~J., {et~al.} 1995, \ssr, 71,
  231

\bibitem[{{Chen} {et~al.}(2020){Chen}, {Kontar}, {Chrysaphi}, {Jeffrey},
  {Gordovskyy}, {Yan}, \& {Tan}}]{2020ApJ...905...43C}
{Chen}, X., {Kontar}, E.~P., {Chrysaphi}, N., {et~al.} 2020, \apj, 905, 43

\bibitem[{{Chrysaphi} {et~al.}(2018){Chrysaphi}, {Kontar}, {Holman}, \&
  {Temmer}}]{2018ApJ...868...79C}
{Chrysaphi}, N., {Kontar}, E.~P., {Holman}, G.~D., \& {Temmer}, M. 2018, \apj,
  868, 79

\bibitem[{{Clarkson} {et~al.}(2023){Clarkson}, {Kontar}, {Vilmer},
  {Gordovskyy}, {Chen}, \& {Chrysaphi}}]{2023ApJ...946...33C}
{Clarkson}, D.~L., {Kontar}, E.~P., {Vilmer}, N., {et~al.} 2023, \apj, 946, 33

\bibitem[{{Dulk} {et~al.}(1984){Dulk}, {Steinberg}, \&
  {Hoang}}]{1984A&A...141...30D}
{Dulk}, G.~A., {Steinberg}, J.~L., \& {Hoang}, S. 1984, \aap, 141, 30

\bibitem[{{Ergun} {et~al.}(1998){Ergun}, {Larson}, {Lin}, {McFadden},
  {Carlson}, {Anderson}, {Muschietti}, {McCarthy}, {Parks}, {Reme}, {Bosqued},
  {D'Uston}, {Sanderson}, {Wenzel}, {Kaiser}, {Lepping}, {Bale}, {Kellogg}, \&
  {Bougeret}}]{1998ApJ...503..435E}
{Ergun}, R.~E., {Larson}, D., {Lin}, R.~P., {et~al.} 1998, \apj, 503, 435

\bibitem[{{Evans} {et~al.}(1973){Evans}, {Fainberg}, \&
  {Stone}}]{1973SoPh...31..501E}
{Evans}, L.~G., {Fainberg}, J., \& {Stone}, R.~G. 1973, \solphys, 31, 501

\bibitem[{{Hewish}(1958)}]{1958MNRAS.118..534H}
{Hewish}, A. 1958, \mnras, 118, 534

\bibitem[{{Jebaraj} {et~al.}(2023){Jebaraj}, {Krasnoselskikh}, {Pulupa},
  {Magdalenic}, \& {Bale}}]{2023ApJ...955L..20J}
{Jebaraj}, I.~C., {Krasnoselskikh}, V., {Pulupa}, M., {Magdalenic}, J., \&
  {Bale}, S.~D. 2023, \apjl, 955, L20

\bibitem[{{Kontar} {et~al.}(2019){Kontar}, {Chen}, {Chrysaphi}, {Jeffrey},
  {Emslie}, {Krupar}, {Maksimovic}, {Gordovskyy}, \&
  {Browning}}]{2019ApJ...884..122K}
{Kontar}, E.~P., {Chen}, X., {Chrysaphi}, N., {et~al.} 2019, \apj, 884, 122

\bibitem[{{Kontar} {et~al.}(2023){Kontar}, {Emslie}, {Clarkson}, {Chen},
  {Chrysaphi}, {Azzollini}, {Jeffrey}, \& {Gordovskyy}}]{2023ApJ...956..112K}
{Kontar}, E.~P., {Emslie}, A.~G., {Clarkson}, D.~L., {et~al.} 2023, \apj, 956,
  112

\bibitem[{{Kontar} {et~al.}(2017){Kontar}, {Yu}, {Kuznetsov}, {Emslie},
  {Alcock}, {Jeffrey}, {Melnik}, {Bian}, \&
  {Subramanian}}]{2017NatCo...8.1515K}
{Kontar}, E.~P., {Yu}, S., {Kuznetsov}, A.~A., {et~al.} 2017, Nature
  Communications, 8, 1515

\bibitem[{{Krupar} {et~al.}(2018){Krupar}, {Maksimovic}, {Kontar}, {Zaslavsky},
  {Santolik}, {Soucek}, {Kruparova}, {Eastwood}, \&
  {Szabo}}]{2018ApJ...857...82K}
{Krupar}, V., {Maksimovic}, M., {Kontar}, E.~P., {et~al.} 2018, \apj, 857, 82

\bibitem[{{Krupar} {et~al.}(2020){Krupar}, {Szabo}, {Maksimovic}, {Kruparova},
  {Kontar}, {Balmaceda}, {Bonnin}, {Bale}, {Pulupa}, {Malaspina}, {Bonnell},
  {Harvey}, {Goetz}, {Dudok de Wit}, {MacDowall}, {Kasper}, {Case}, {Korreck},
  {Larson}, {Livi}, {Stevens}, {Whittlesey}, \&
  {Hegedus}}]{2020ApJS..246...57K}
{Krupar}, V., {Szabo}, A., {Maksimovic}, M., {et~al.} 2020, \apjs, 246, 57

\bibitem[{{Kuznetsov} {et~al.}(2020){Kuznetsov}, {Chrysaphi}, {Kontar}, \&
  {Motorina}}]{2020ApJ...898...94K}
{Kuznetsov}, A.~A., {Chrysaphi}, N., {Kontar}, E.~P., \& {Motorina}, G. 2020,
  \apj, 898, 94

\bibitem[{{Li} {et~al.}(2012){Li}, {Cairns}, \&
  {Robinson}}]{2012SoPh..279..173L}
{Li}, B., {Cairns}, I.~H., \& {Robinson}, P.~A. 2012, \solphys, 279, 173

\bibitem[{{Lin}(1985)}]{1985SoPh..100..537L}
{Lin}, R.~P. 1985, \solphys, 100, 537

\bibitem[{{Maksimovic} {et~al.}(2020){Maksimovic}, {Bale}, {Chust},
  {Khotyaintsev}, {Krasnoselskikh}, {Kretzschmar}, {Plettemeier}, {Rucker},
  {Sou{\v{c}}ek}, {Steller}, {{\v{S}}tver{\'a}k}, {Tr{\'a}vn{\'\i}{\v{c}}ek},
  {Vaivads}, {Chaintreuil}, {Dekkali}, {Alexandrova}, {Astier}, {Barbary},
  {B{\'e}rard}, {Bonnin}, {Boughedada}, {Cecconi}, {Chapron}, {Chariet},
  {Collin}, {de Conchy}, {Dias}, {Gu{\'e}guen}, {Lamy}, {Leray}, {Lion},
  {Malac-Allain}, {Matteini}, {Nguyen}, {Pantellini}, {Parisot}, {Plasson},
  {Thijs}, {Vecchio}, {Fratter}, {Bellouard}, {Lorf{\`e}vre}, {Danto},
  {Julien}, {Guilhem}, {Fiachetti}, {Sanisidro}, {Laffaye}, {Gonzalez},
  {Pontet}, {Qu{\'e}ruel}, {Jannet}, {Fergeau}, {Brochot}, {Cassam-Chenai},
  {Dudok de Wit}, {Timofeeva}, {Vincent}, {Agrapart}, {Delory}, {Turin},
  {Jeandet}, {Leroy}, {Pellion}, {Bouzid}, {Katra}, {Piberne}, {Recart},
  {Santol{\'\i}k}, {Kolma{\v{s}}ov{\'a}}, {Krupa{\v{r}}},
  {Krupa{\v{r}}ov{\'a}}, {P{\'\i}{\v{s}}a}, {Uhl{\'\i}{\v{r}}}, {L{\'a}n},
  {Ba{\v{s}}e}, {Ahl{\`e}n}, {Andr{\'e}}, {Bylander}, {Cripps}, {Cully},
  {Eriksson}, {Jansson}, {Johansson}, {Karlsson}, {Puccio},
  {B{\v{r}}{\'\i}nek}, {{\"O}ttacher}, {Panchenko}, {Berthomier}, {Goetz},
  {Hellinger}, {Horbury}, {Issautier}, {Kontar}, {Krucker}, {Le Contel},
  {Louarn}, {Martinovi{\'c}}, {Owen}, {Retino}, {Rodr{\'\i}guez-Pacheco},
  {Sahraoui}, {Wimmer-Schweingruber}, {Zaslavsky}, \&
  {Zouganelis}}]{2020A&A...642A..12M}
{Maksimovic}, M., {Bale}, S.~D., {Chust}, T., {et~al.} 2020, \aap, 642, A12

\bibitem[{{Melrose}(2017)}]{2017RvMPP...1....5M}
{Melrose}, D.~B. 2017, Reviews of Modern Plasma Physics, 1, 5

\bibitem[{{M{\"u}ller} {et~al.}(2020){M{\"u}ller}, {St. Cyr}, {Zouganelis},
  {Gilbert}, {Marsden}, {Nieves-Chinchilla}, {Antonucci}, {Auch{\`e}re},
  {Berghmans}, {Horbury}, {Howard}, {Krucker}, {Maksimovic}, {Owen}, {Rochus},
  {Rodriguez-Pacheco}, {Romoli}, {Solanki}, {Bruno}, {Carlsson}, {Fludra},
  {Harra}, {Hassler}, {Livi}, {Louarn}, {Peter}, {Sch{\"u}hle}, {Teriaca}, {del
  Toro Iniesta}, {Wimmer-Schweingruber}, {Marsch}, {Velli}, {De Groof},
  {Walsh}, \& {Williams}}]{2020A&A...642A...1M}
{M{\"u}ller}, D., {St. Cyr}, O.~C., {Zouganelis}, I., {et~al.} 2020, \aap, 642,
  A1

\bibitem[{{Musset} {et~al.}(2021){Musset}, {Maksimovic}, {Kontar}, {Krupar},
  {Chrysaphi}, {Bonnin}, {Vecchio}, {Cecconi}, {Zaslavsky}, {Issautier},
  {Bale}, \& {Pulupa}}]{2021A&A...656A..34M}
{Musset}, S., {Maksimovic}, M., {Kontar}, E., {et~al.} 2021, \aap, 656, A34

\bibitem[{{Parker}(1958)}]{1958ApJ...128..664P}
{Parker}, E.~N. 1958, \apj, 128, 664

\bibitem[{{Poquerusse}(1977)}]{1977A&A....56..251P}
{Poquerusse}, M. 1977, \aap, 56, 251

\bibitem[{{Pulupa} {et~al.}(2020){Pulupa}, {Bale}, {Badman}, {Bonnell}, {Case},
  {de Wit}, {Goetz}, {Harvey}, {Hegedus}, {Kasper}, {Korreck},
  {Krasnoselskikh}, {Larson}, {Lecacheux}, {Livi}, {MacDowall}, {Maksimovic},
  {Malaspina}, {Mart{\'\i}nez Oliveros}, {Meyer-Vernet}, {Moncuquet},
  {Stevens}, \& {Whittlesey}}]{2020ApJS..246...49P}
{Pulupa}, M., {Bale}, S.~D., {Badman}, S.~T., {et~al.} 2020, \apjs, 246, 49

\bibitem[{{Pulupa} {et~al.}(2017){Pulupa}, {Bale}, {Bonnell}, {Bowen},
  {Carruth}, {Goetz}, {Gordon}, {Harvey}, {Maksimovic},
  {Mart{\'\i}nez-Oliveros}, {Moncuquet}, {Saint-Hilaire}, {Seitz}, \&
  {Sundkvist}}]{2017JGRA..122.2836P}
{Pulupa}, M., {Bale}, S.~D., {Bonnell}, J.~W., {et~al.} 2017, Journal of
  Geophysical Research (Space Physics), 122, 2836

\bibitem[{{Ratcliffe} {et~al.}(2014){Ratcliffe}, {Kontar}, \&
  {Reid}}]{2014A&A...572A.111R}
{Ratcliffe}, H., {Kontar}, E.~P., \& {Reid}, H.~A.~S. 2014, \aap, 572, A111

\bibitem[{{Reid} \& {Kontar}(2018{\natexlab{a}})}]{2018A&A...614A..69R}
{Reid}, H. A.~S. \& {Kontar}, E.~P. 2018{\natexlab{a}}, \aap, 614, A69

\bibitem[{{Reid} \& {Kontar}(2018{\natexlab{b}})}]{2018ApJ...867..158R}
{Reid}, H. A.~S. \& {Kontar}, E.~P. 2018{\natexlab{b}}, \apj, 867, 158

\bibitem[{{Reid} \& {Kontar}(2021)}]{2021NatAs...5..796R}
{Reid}, H. A.~S. \& {Kontar}, E.~P. 2021, Nature Astronomy, 5, 796

\bibitem[{{Steinberg} {et~al.}(1971){Steinberg}, {Aubier-Giraud}, {Leblanc}, \&
  {Boischot}}]{1971A&A....10..362S}
{Steinberg}, J.~L., {Aubier-Giraud}, M., {Leblanc}, Y., \& {Boischot}, A. 1971,
  \aap, 10, 362

\bibitem[{{Vecchio} {et~al.}(2021){Vecchio}, {Maksimovic}, {Krupar}, {Bonnin},
  {Zaslavsky}, {Astier}, {Dekkali}, {Cecconi}, {Bale}, {Chust}, {Guilhem},
  {Khotyaintsev}, {Krasnoselskikh}, {Kretzschmar}, {Lorf{\`e}vre},
  {Plettemeier}, {Sou{\v{c}}ek}, {Steller}, {{\v{S}}tver{\'a}k},
  {Tr{\'a}vn{\'\i}{\v{c}}ek}, \& {Vaivads}}]{2021A&A...656A..33V}
{Vecchio}, A., {Maksimovic}, M., {Krupar}, V., {et~al.} 2021, \aap, 656, A33

\end{thebibliography}

%-----------------------------------------------------------------------
\begin{appendix}

%=================================
\section{Event selection criteria} \label{sec:selection_criteria}
\begin{figure}[ht!]
    \centering
    \includegraphics[width=0.4\textwidth, keepaspectratio=true]{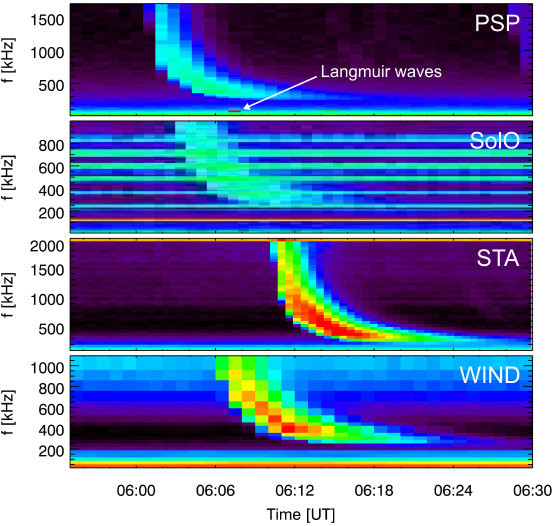}
    \caption{Dynamic spectra of a Type III burst observed by four spacecraft on 22-Aug-2021.  The spacecraft used to record each dynamic spectrum is indicated in each panel, and the observed Langmuir waves are also annotated.  Frequencies in SolO spectra suffering from perturbations were omitted during analysis.  Unprocessed Level 2 data, as made publicly available, are used in this figure.  The recorded signals are not calibrated, so the colours are arbitrary and do not represent the true relative strength of the signal measured by each of the spacecraft.
    }
    \label{fig:multi-spacectaft_observation}
\end{figure}

This study focuses on Type III bursts as they are the most frequently-observed and intense radio bursts, spanning large ranges of frequencies, and their corresponding Langmuir waves are often recorded in situ.
Multi-vantage observations of interplanetary Type III bursts were compared using data from 4 different spacecraft (as shown in Fig.~\ref{fig:multi-spacectaft_observation}):
\begin{enumerate}[label=(\roman*)]
    \item the Thermal Noise Receiver (TNR) of the Radio and Plasma Waves \citep[RPW;][]{2020A&A...642A..12M} instrument onboard Solar Orbiter \citep[SolO;][]{2020A&A...642A...1M},
    \item the Low-Frequency Receiver (LFR) of the Radio Frequency Spectrometer \citep[RFS;][]{2017JGRA..122.2836P} on the Electromagnetic Fields Investigation \citep[FIELDS;][]{2016SSRv..204...49B} onboard the Parker Solar Probe (PSP),
    \item the High Frequency Receiver (HFR) of the Radio and Plasma Wave Investigation \citep[SWAVES;][]{2008SSRv..136..487B} onboard the Solar TErrestrial RElations Observatory (STEREO-A; or STA), and
    \item the RAD1 receiver of the Radio and Plasma Wave Investigation \citep[WAVES;][]{1995SSRv...71..231B}) onboard WIND.
\end{enumerate}
The Type III events used for analysis were selected based on the following criteria:
\begin{enumerate}
    \item The same Type III burst is observed by at least 3 spacecraft at various longitudes, allowing for a multi-vantage analysis.
    \item Langmuir waves associated to the Type III bursts (i.e. they are temporally and spatially close) are recorded in situ by either the SolO or PSP spacecraft.
    This criterion allows us to approximate the location of the source in the heliosphere, since the spacecraft that is measuring the Langmuir waves in situ can be considered as traversing through the radio excitation region \citep{1985SoPh..100..537L, 1998ApJ...503..435E, 2020ApJS..246...49P}.
    \item Isolated Type III bursts are observed, such that no other bursts contribute to the recorded properties of the emissions of interest.
    \item No radio storm is present.  Contrary cases may have been accepted if the selected Type III burst is sufficiently more intense than any surrounding (background) emissions.
    \item No multiple injections can be (at least clearly) distinguished, eliminating the possibility that the observed radio emissions result from multiple Type III bursts (i.e. multiple electron beam excitations) that are interwoven together.
\end{enumerate}

%------------------------------------FIGURE-------------------------------------
%----------------------------------END FIGURE-----------------------------------

Data from late 2018 (i.e. since the first measurements by PSP/FIELDS) until the end of June 2022 was examined.
Following the elimination of candidates that did not meet all these criteria (or doubts about some of them existed), only nine Type III events were selected for analysis in this study (listed in Table~\ref{tab:events_list}).
In all cases, Level 2 spacecraft data was used, where the recorded signal is given in $\mathrm{V^2/Hz}$.  For WIND, the 60~s-averaged data was utilised.
For the analysis of each spacecraft's observation, measurements from the antenna configuration that recorded the strongest Type III signal were preferred.

%------------------------------------TABLE--------------------------------------
\begin{table*}[ht!]
    \renewcommand{\arraystretch}{1.3}
    \caption{
    List of analysed Type III events.
    }
    \centering
    \begin{tabular}{|c||c|c|c|c|c|}
        \hline
         Date & $\mathrm{t_{\_LW}}$ [UT] (s/c) & $\mathrm{f_{\_PSP}}$ [kHz] & $\mathrm{f_{\_SOLO}}$ [kHz] & $\mathrm{f_{\_STA}}$ [kHz] & $\mathrm{f_{\_WIND}}$ [kHz] \\
         \hline
         18/11/2020 & 19:10 (SolO) & 328.12 & 331.31 & 325 & 332 \\
         \hline
         16/07/2021 & 09:20 (PSP) & 292.97 & 290.93 & 275 & 292 \\
         \hline
         22/08/2021 & 06:10 (PSP) & 276.56 & 266.79 & 275 & 256 \\
         \hline
         30/08/2021 & 01:40 (PSP) & 366.8 & 377.29 & 375 & 376 \\
         \hline
         06/10/2021 & 04:30 (SolO) & (...) & 377.29 & 375 & 376 \\
         \hline
         05/02/2022 & 10:30 (PSP) & (...) & 290.93 & 275 & 292 \\
         \hline
         10/05/2022 & 04:00 (SolO) & 292.97 & 290.93 & 275 & 292 \\
         \hline
         16/05/2022 & 00:40 (PSP) & 328.12 & 331.31 & (...) & 332 \\
         \hline
         18/05/2022 & 00:50 (PSP) & 544.92 & 533.57 & 525 & 548 \\
         \hline
    \end{tabular}
    \tablefoot{
    $\mathrm{t_{\_LW}}$ is the approximate observation time of the Langmuir waves associated to the Type III bursts, given with the spacecraft (s/c) that recorded these Langmuir waves.  The frequency at which each event was analysed for the purposes of examining the decay and rise times as a function of the observer's location is given for each of the four spacecraft ($\mathrm{f_{\_PSP}}$, $\mathrm{f_{\_SOLO}}$, $\mathrm{f_{\_STA}}$, and $\mathrm{f_{\_WIND}}$).  (...) indicates the lack of observation or lack of fitting output (due to a faint signal).
    }
    \label{tab:events_list}
\end{table*}
%----------------------------------END TABLE------------------------------------

No limits were specified on the angular separation between the spacecraft for the selection of these events.  Here, it should also be emphasised that the only meaningful angular separation for such analyses is that calculated with respect to the radio source and in the frame of the source and detector (where the source is considered to be co-spatial with a spacecraft).  This is the angular separation calculated in this study, as discussed in Appendix~\ref{sec:angle_calculation}.
Moreover, the heliocentric distance of the spacecraft was not considered as a criterion either.  It should be clarified that the heliocentric distance of the spacecraft alone is not a meaningful quantity for such studies.  What is physically meaningful is the (Euclidean) distance of the spacecraft from the radio source itself.  However, as illustrated in Appendix~\ref{sec:decay_rise_vs_distance}, the Euclidean distance of the spacecraft does not appear to influence the measurements.

Interplanetary Type III bursts have bandwidths spanning a large range of frequencies, but for the purposes of examining the rise and decay times as a function of the observer's separation from the source, lower frequencies are preferred.  This approach was followed in order to ensure that the analysed frequency was not far off from that at which the Langmuir waves were observed, allowing for the assumption that the position of the spacecraft recording the Langmuir waves in situ is analogous to the source position used for the estimation of the angular and Euclidean separations (Appendix~\ref{sec:angle_calculation}).  However, light curves with frequencies virtually the same as those of the Langmuir waves (and presumably very close to the source) were avoided, to make sure that the shape of the examined time profile is fully-evolved (see Sect.~\ref{sec:conclusion}).
A benefit of selecting lower-frequency light curves for the analysis is that they are broader than their higher-frequency counterparts.  This means that for an instrument of given temporal resolution, lower-frequency light curves comprise of more data points, and consequently their shape is described in a more complete manner.

It should be emphasised that the fifth criterion applied for the event selection does not impact the analysis on the rise and decay time measurements as a function of the observer's location (Sect.~\ref{sec:decay_rise_vs_angle} and Appendix~\ref{sec:decay_rise_vs_distance}).  Specifically, the possibility of multiple injections of electron beams convolving to form the Type III bursts analysed would not interference with any physical mechanism that may lead to the rise and decay time measurements depending on the (angular or Euclidean) position of the observer.  However, the entanglement of several electron beams forming the interplanetary Type III bursts could impact the exact values of the calculated rise and decay times, and thus the absolute value of their ratio (Sect.~\ref{sec:rise_decay_ratio}).  Whether any interplanetary Type III burst can result from a single electron beam remains an open question.  Nevertheless, we only analyse events that showed no clear signs of multiple injections at higher frequencies, which, along with the second criterion, are the predominant factors limiting our study to nine events. 

Even though some of the analysed bursts clearly result from fundamental emissions (given the relation to their Langmuir waves), no distinction was made between fundamental and harmonic emissions.  We note that it is theoretically possible that fundamental and harmonic emissions are simultaneously present in interplanetary Type III bursts \citep[e.g.][]{1984A&A...141...30D, 2023ApJ...955L..20J}.
Such possibility, though, does not affect our analysis on the dependence of the decay and rise times on the observer's position (Sect.~\ref{sec:decay_rise_vs_angle} and Appendix~\ref{sec:decay_rise_vs_distance}).  It could, however, affect the absolute values of the estimated rise and decay times, and thus their exact ratio.  Nonetheless, we find that the rise-to-decay time ratio of both fundamental and harmonic emissions shows a frequency independence, as discussed in Sect.~\ref{sec:rise_decay_ratio}, meaning that our conclusions are unaffected by this possibility.

%=================================
\section{Angular separation calculation} \label{sec:angle_calculation}
The 3D positions of the spacecraft are considered, denoted by $R = (x, y, z)$ in Heliocentric Cartesian coordinates, such that $\vec{R}=\vec{x}+\vec{y}+\vec{z}$ is the Sun-spacecraft vector.
The radio source position is approximated as the position of the spacecraft observing the Langmuir waves in situ (here at $R_1$), such that the spacecraft at $R_1$ defines the Sun-source vector.
The angular separation $\theta$ between the source/spacecraft at $R_1$ and a spacecraft at $R_2$ is measured from the Sun-source axis to the vector $\vec{R_{12}}$ connecting the two spacecraft (whose magnitude is their Euclidean distance; see Fig.~\ref{fig:angle_cartoon}).
Therefore, the spacecraft at $R_1$ is on the $\theta=0\degr$ axis.
The angular separation is given as
\begin{equation} \label{eqn:ang_separ}
    \theta = \cos^{-1} \left( \frac{\vec{R_1} \cdot \vec{R_{12}}}{\left|\vec{R_1}\right| \left|\vec{R_{12}}\right|} \right) \, ,
\end{equation}
where $\theta$ is in the plane of the two spacecraft and is given in radians (and then transformed into degrees), and $\vec{R_{12}}=(\vec{x}_2-\vec{x}_1, \vec{y}_2-\vec{y}_1, \vec{z}_2-\vec{z}_1)$.

It is emphasised that the angle of interest is not that between the $\vec{R_1}$ and $\vec{R_2}$ vectors, which would correspond to the heliocentric angle $\theta_H$ shown in Fig.~\ref{fig:angle_cartoon}.  Instead, $\theta$ calculated between vectors $\vec{R_1}$ and $\vec{R_{12}}$, as given by Eq.~(\ref{eqn:ang_separ}), represents the meaningful angular separation for this study.
This is because the interest is in the angular displacement of a given detector from the maximum of the (scattered) radio-photon beam (i.e. the ``beam axis'').  Due to the presence of anisotropic density fluctuations which are aligned along the magnetic field, radio emissions are directional.  In other words, the photon emissions are not isotropic, and thus observed radio source properties can vary with the angular separation from this beam axis.
For a certain period following the excitation of radio photons from the source, the photons' beam axis is aligned with the magnetic field -- which can be assumed (for interplanetary emissions) to have a Parker spiral configuration \citep{1958ApJ...128..664P} -- since density fluctuations are aligned with it too.  Once far enough from the source where scattering effects weaken, the beam axis is no longer aligned with the magnetic field, but follows a more tangential trajectory (see Appendix~\ref{sec:simulations} for more details).
Therefore, we approximate the beam axis as being aligned with the Sun-source axis, and consequently measure the angular separation $\theta$ from it.  It should be emphasised that the approximation of a fixed beam axis along the Sun-source axis is a simplifying one.  However, given that the exact 3D orientation of the magnetic field cannot be evaluated at any given location and time during the emissions (depending on the frequency and time examined), it is a satisfactory assumption that introduces as many uncertainties as any other reasonable assumptions would (like using the measurements at the location of the spacecraft, or using approximations of the Parker spiral configuration for each event), while maintaining meaningful angular-separation estimations.

%=================================
\section{Reliability of the fitting function} \label{sec:func_reliability}

%------------------------------------FIGURE-------------------------------------
\begin{figure*}[ht!]
    \centering
    \includegraphics[width=0.45\textwidth, keepaspectratio=true]{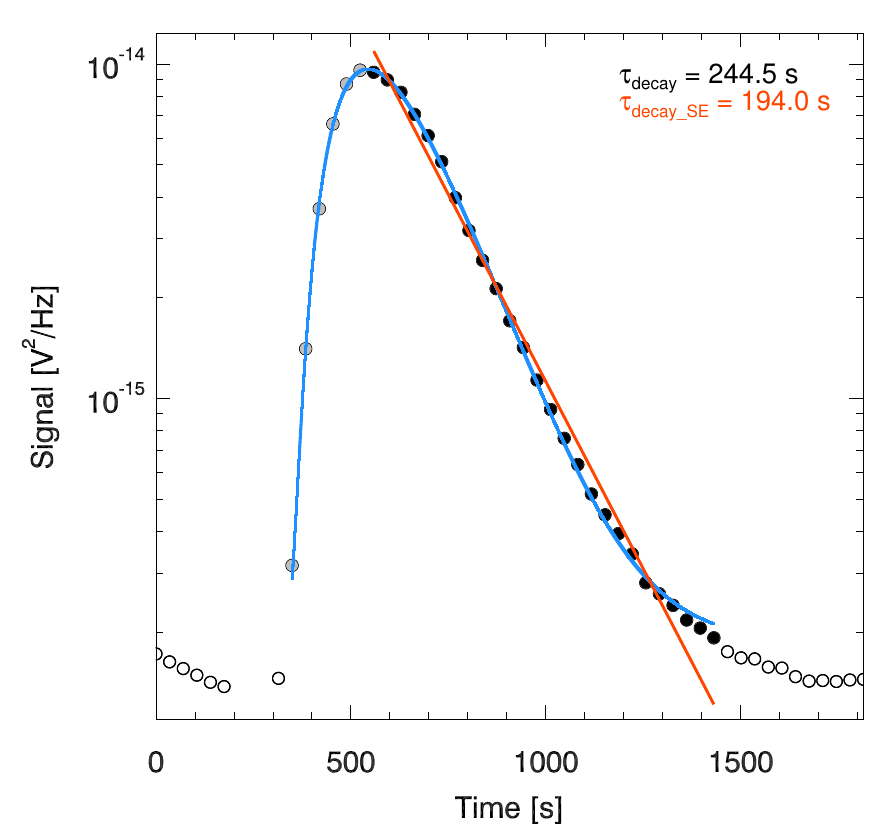}~
    \hspace{-1em}
    \includegraphics[width=0.45\textwidth, keepaspectratio=true]{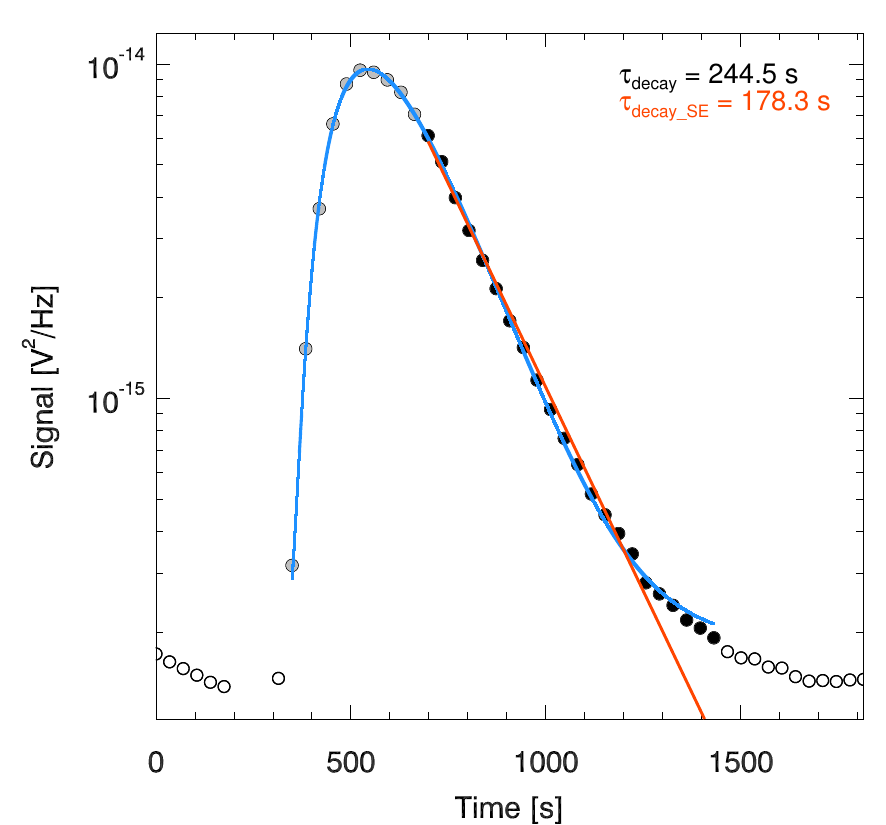}
    \includegraphics[width=0.45\textwidth, keepaspectratio=true]{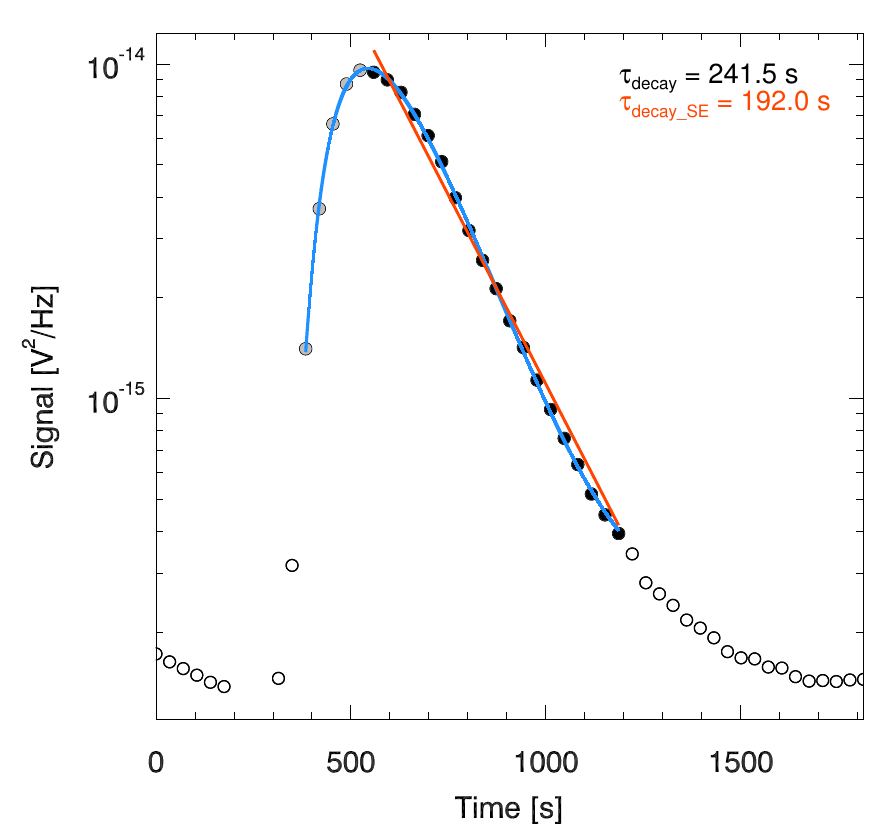}~
    \hspace{-1em}
    \includegraphics[width=0.45\textwidth, keepaspectratio=true]{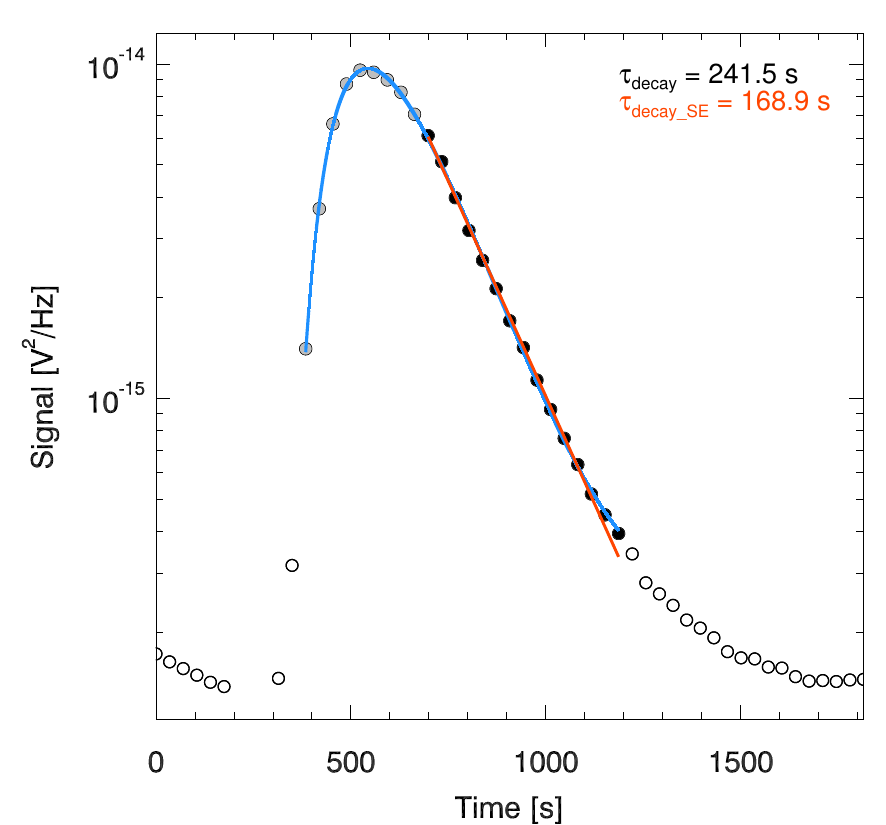}
    \caption{
    Comparisons of the proposed function (Eq.~(\ref{eqn:fit_function})) and the single-exponential function (Eq.~(\ref{eqn:single_exp})), using 4 different combinations of user-defined fit conditions (detailed in the text).
    The corresponding decay times obtained from the proposed and single-exponential (SE) fit are indicated in the legend.
    Top row: Fitting only the signals $\ge$2\% of the peak amplitude.
    Bottom row: Fitting only the signals $\ge$4\% of the peak amplitude.
    Left column: Excluding the peak-amplitude data point from the single-exponential fit.
    Right column:  Excluding the first 5 data points after the peak time from the single-exponential fit, ignoring the initial part of the decay phase where it appears to be non-exponential.
    Grey, filled circles indicate the data points considered when fitting using the proposed function (blue curve).  Black, filled circles indicate the data points considered for the single-exponential fit (orange line).
    The legend of each panel depicts the estimated decay time from both the proposed fit (in black) and the single exponential fit (in orange).
    The data was recorded by STA at 325~kHz around 17:50~UT on 18-Nov-2020, as shown in Fig.~\ref{fig:fit_function}a.
    }
    \label{fig:fit_stability_comparisons}
\end{figure*}
%----------------------------------END FIGURE-----------------------------------

%------------------------------------FIGURE-------------------------------------
\begin{figure}[ht!]
    \centering
    \includegraphics[width=0.5\textwidth, keepaspectratio=true]{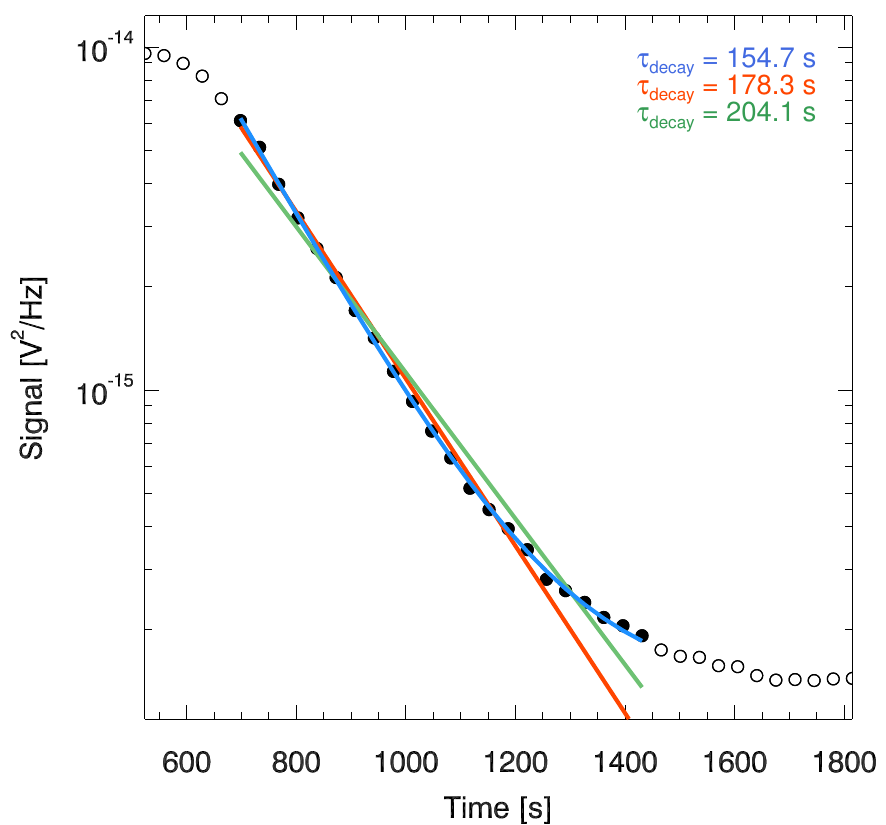}~
    \caption{
    Comparisons of single-exponential fit outputs for 3 differing user choices.
    The orange fit is obtained by fitting the data using Eq.~(\ref{eqn:single_exp}), as in the top right panel of Fig.~\ref{fig:fit_stability_comparisons}.
    The green fit is obtained by linearising the data before fitting it with the linearised form of Eq.~(\ref{eqn:single_exp}).
    The blue curve was obtained by fitting the data with a variant of Eq.~(\ref{eqn:single_exp}) that includes the addition of a constant offset as a fit parameter.
    The corresponding estimated decay times are indicated in the legend, highlighting the significant variation arising from these user choices.
    The data was recorded by STA at 325~kHz around 17:50~UT on 18-Nov-2020.
    }
    \label{fig:sing_exp_fit_comparisons}
\end{figure}
%----------------------------------END FIGURE-----------------------------------

It has been argued that the initial part of the decay phase -- where the signal is sometimes observed as non-exponential -- must be excluded from fits calculating the decay time \citep[e.g.][]{1972A&A....19..343A, 1975SoPh...45..459B}.
However, it should be emphasised that the exact turning -- if observed at all -- depends heavily on the temporal resolution of the measurement.  This can lead to variations between different studies depending on the quality of the data under examination.
Moreover, the higher-amplitude measurements are the ones that can be trusted the most, since they have larger signal-to-noise ratios.
An argument for this approach is the possibility that the initial part of the decay phase may be non-exponential due to some excited photons arriving at the detector after the peak time, contributing to the decay-phase measurements \citep[e.g.][]{1972A&A....19..343A}.
Notably, for this approach to be physically meaningful, one has to assume that any part which is non-exponential is purely the contribution of the excitation process itself \citep{1972A&A....19..343A}, and that the mechanism(s) contributing to the decay of the radio emissions must (all) be purely exponential.
However, there is currently no evidence to substantiate both of these assumptions.
It has been demonstrated that radio-wave scattering on small-scale density fluctuations dominates the observed decay of radio signals, and such scattering effects lead to an exponential decay \citep[e.g.][]{2018ApJ...857...82K, 2019ApJ...873...33B, 2019ApJ...884..122K, 2020ApJ...905...43C}, but the convolution of several processes cannot be excluded.
Therefore, there is no unambiguous physical motivation for excluding parts of the decay phase from fits, besides those that fall below the background-noise level (Appendix~\ref{sec:bg_estimation}).

The fitting function we propose in Eq.~(\ref{eqn:fit_function}) is advantageous to the commonly-used single-exponential fit to the decay phase, as it characterises the entirety of the light curve and provides a simultaneous estimation of the decay time, the rise time, and the peak flux.  However, the proposed function can also provide improved estimations of the decay time, as it is more reliable than the single-exponential fit and its outputs are much less dependent on user choices.
Using data recorded by a spacecraft, it will be shown here that large variations in the decay times estimated using single-exponential fits can occur even within a single light curve, depending on the user choices.  This indicates the large (unaccounted for) uncertainties that may be introduced when comparing single-exponential outputs between different frequencies of a single event, different instruments, different radio burst events, or even different studies.
The two main user choices invoked in single-exponential fits that can impact the obtained values are:
\begin{enumerate*}[label=(\alph*)]
    \item the starting point (in the decay phase) to be considered for the fit, and
    \item the minimum signal to be considered.
\end{enumerate*}
The intrinsic errors likely induced by such user choices are not reflected in any error estimations of measurements obtained from single-exponential fits.
Crucially, the chosen decay-phase starting point is only affecting the outputs of single-exponential fits (Eq.~(\ref{eqn:single_exp})), since our proposed function (Eq.~(\ref{eqn:fit_function})) does not neglect the data measurements prior to the peak amplitude, thus eliminating one of these user-induced errors.

The reliability of the calculated decay times from the two fitting functions are compared using a single light curve in Fig.~\ref{fig:fit_stability_comparisons}, where higher-amplitude signals were weighted more heavily.  Four different combinations of user-defined conditions are depicted, by considering two different minimum-signal thresholds (2\% and 4\% of the peak-signal amplitude) and two different starting points for the fit after the peak time.
In the first case of single-exponential fits, all data after (and excluding) the peak-time measurement were considered, whereas in the second case, the first 5 measurements after the peak time were excluded, ignoring the initial part of the decay phase that appears to be non-exponential.
As can be seen in Fig.~\ref{fig:fit_stability_comparisons}, all four combinations result in reasonable single-exponential fits, but the corresponding decay-time estimations can vary significantly, highlighting the sensitivity of the single-exponential fit to these user-defined conditions, and that it is prone to resolution-dependent errors.  By contrast, the variation in the decay times obtained by the function we propose is much more limited, illustrating its reliability and consistency.

Fitting the same light curve (in the same way) and choosing to exclude just the peak-amplitude measurement from the dataset or choosing to exclude the first five (non-exponential) measurements, yields a decay-time variation of $\sim$8.8\% ($\sim$13.7\%) assuming a minimum-signal threshold of 2\% (4\%).
The choice of minimum-signal threshold itself (2\% or 4\%) affects the obtained decay time by $\sim$5.6\% when the first 5 measurements are excluded from the fit.  In comparison, for the same two minimum-signal thresholds, our proposed function leads to a decay-time estimation difference of only $\sim$1.2\%.  Therefore, our function proves to be more stable than the single-exponential fit.

These comparisons demonstrate that identical datasets can result in significantly-different decay-time estimations, depending on the way the single-exponential fit is applied.  This further highlights the likely intrinsic errors in decay-time estimations based on single-exponential fits, and the potential scatter affecting comparisons of decay times of the same event observed by different instruments, which likely have different resolutions and are affected by varying background-noise levels.

There are two other user choices that can significantly impact the decay times obtained from a single-exponential fit, as demonstrated in Fig.~\ref{fig:sing_exp_fit_comparisons}.
For all the fits shown here, the same (weighted) fitting algorithm was used, with the same minimum-signal threshold (2\%), and excluding the first 5 data points of the decay phase.
The first user choice having an impact on the results is whether the data is fit with an exponential function, or whether the data is first linearised (to accentuate the exponential parts of the curve) and then fitted \citep[e.g.,][]{1972A&A....19..343A, 1973SoPh...31..501E, 1975SoPh...45..459B, 1977A&A....56..251P}.
However, as illustrated in Fig.~\ref{fig:sing_exp_fit_comparisons}, linearising the data before fitting it can result in worst fits, due to the nature of the minimisation of fitting algorithms themselves.
To obtain the green fit, data was linearised prior to fitting with the linearised form of Eq.~(\ref{eqn:single_exp}), $\ln{(I_d)} = \ln{(I_{max})} - \frac{1}{\tau_d}(t_d-t_0)$, whereas the orange fit was obtained by directly fitting the data using Eq.~(\ref{eqn:single_exp}), resulting in a better fit and a decay-time variation of $\sim$14.5\%.
The second user choice is the addition of a constant offset as a fit parameter in Eq.~(\ref{eqn:single_exp}), which can also significantly affect the obtained decay time from single-exponential fits, and is frequently omitted from fits \citep[e.g.,][]{2018A&A...614A..69R, 2018ApJ...857...82K, 2020ApJS..246...57K, 2021A&A...656A..33V}.
However, as indicated in Fig.~\ref{fig:sing_exp_fit_comparisons}, its addition can improve the obtained fit (blue curve), and leads to yet another decay time value, which differs by $\sim$15.3\% compared to the equivalent fit without the constant offset (orange line).
These further illustrate the potential variability and uncertainty in meaningfully comparing decay-time measurements obtained by different studies that have applied a single-exponential fit.

%=================================
\section{Background estimation} \label{sec:bg_estimation}
Solar spectra depicting radio burst emissions also include other radio contributions that affect the intrinsic signal from the radio burst sources.  These contributions include instrumental noise from the spacecraft and instrument itself, radio emissions from the galaxy (galactic background), the local Quasi-Thermal Noise, other solar radio emissions that are interfering with the signal of interest, or sometimes even Jovian emissions \citep[e.g.][]{2020A&A...642A..12M}.  The aggregate contribution of noise and interference defines a base level of recorded emissions, known as the background signal, which varies with time and is frequency dependent.  This background is often removed prior to analysis as it distorts the true amplitude of the recorded radio burst emissions.

Time windows including the isolated Type III bursts analysed (see Appendix~\ref{sec:selection_criteria}) were manually selected for the background level estimation, such that a sufficient ``quiet'' period before and after the Type III burst is present, and no other intense radio emissions contributed to the signal of interest.  The median value of the signal amplitudes comprising the selected window was defined as the constant background level for the specific light curve.  Given its frequency dependence, the background value was estimated for each individual time profile analysed.

%=================================
\section{Prediction of the radio-wave propagation simulations} \label{sec:simulations}
We utilise state-of-the-art 3D ray-tracing simulations of radio-wave propagation effects developed by \cite{2019ApJ...884..122K} that take scattering, refraction, absorption, as well as anisotropic density fluctuations into account, in order to simulate the observed radio source properties, as has been successfully done in several other studies \citep{2019ApJ...884..122K, 2020ApJ...905...43C, 2020ApJ...898...94K, 2021A&A...656A..34M}. 
The configuration of the simulated magnetic field was upgraded from radial to that of the Parker spiral.  The (anisotropic) density fluctuations are aligned with the magnetic field, causing the overall trajectory of radio photons (the beam axis) to be roughly aligned with the magnetic field too (in this case the Parker spiral).  This is true until a distance farther from the source where scattering weakens, specifically, until the point where the density fluctuations no longer define the directivity of the radio waves.  After this distance, the photon beam axis deviates away from the magnetic field (in a more-or-less tangential manner).  The simulations are consistent with and reproduce the observations where a shift of the photon beam axis away from the magnetic field direction was identified \citep[e.g.][]{2008A&A...489..419B}.  Therefore, as expected, simulation outputs are more realistic when the Parker spiral configuration is assumed in place of the radial one, and thus adopted for this analysis.

The simulated radio source properties used in this study represent those that detectors at different longitudes would observe at 1 au.  Given that -- by default -- the simulations define angular separations with respect to the Sun-Earth axis \citep{2019ApJ...884..122K}, the outputs were re-calculated to represent angle $\theta$ used for the analysis of the observations, as depicted in Fig.~\ref{fig:angle_cartoon}.
Radio waves with $f \approx 250$~kHz were simulated (using $\sim 10^6$ photons), a frequency which corresponds to an intrinsic source location at $\sim$26~$\mathrm{R_\sun}$ ($\approx 0.12$~au), assuming fundamental emissions and a ratio of observed frequency to local (electron) plasma frequency $f/f_{pe} = 1.1$.
The assumed scattering strength is $\varepsilon = 0.8$ and the level of anisotropy (a ratio of the perpendicular to the parallel scattering strength) is $\alpha=0.3$, in-line with the input parameters used by \cite{2019ApJ...884..122K} to successfully reproduce the (average) properties of a combination of Type III burst observations.

%------------------------------------FIGURE-------------------------------------
\begin{figure}[ht!]
    \centering
    \includegraphics[width=0.45\textwidth, keepaspectratio=true]{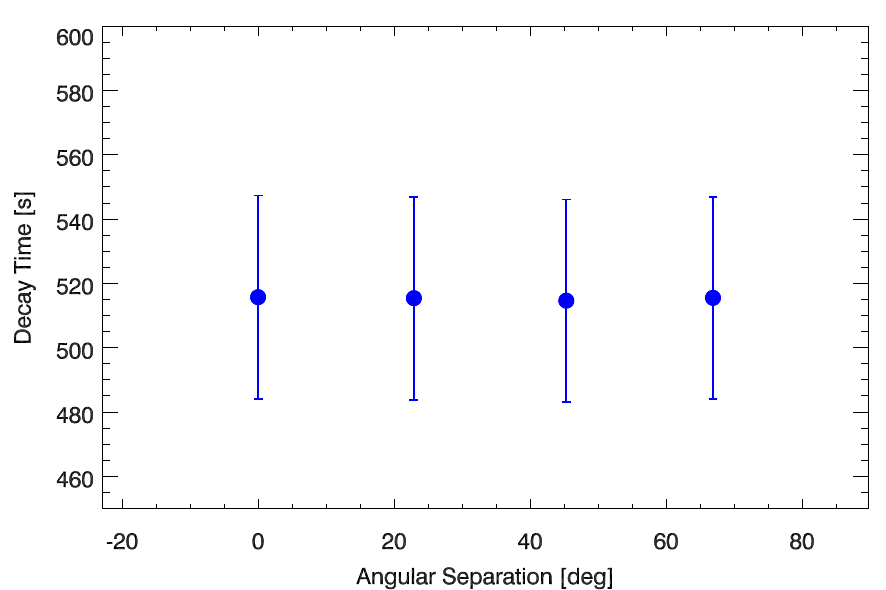}
    \includegraphics[width=0.45\textwidth, keepaspectratio=true]{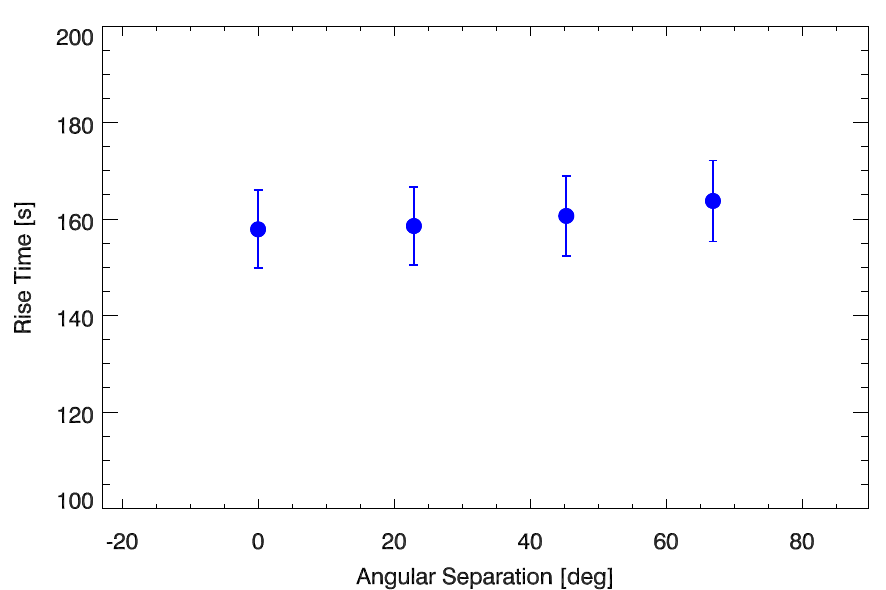}
    \caption{
    Simulated decay time (top) and rise time (bottom) as a function of angular separation, and their associated errors.
    The observers are located at the heliocentric distance of 1~au at varying angles from the source.
    Fundamental emissions with a frequency of $\sim250$~kHz were assumed, as well as a level of scattering $\varepsilon=0.8$ and a level of anisotropy $\alpha=0.3$. 
    }
    \label{fig:simulations}
\end{figure}
%----------------------------------END FIGURE-----------------------------------

The simulation predictions are depicted in Fig.~\ref{fig:simulations}.
The decay and rise times were calculated simultaneously by fitting the entire simulated light curves with the proposed function of Eq.~(\ref{eqn:fit_function}).  The uncertainties on these decay and rise times were estimated by varying the optimised fit parameters by $\pm 10$\%, as done for the observational data (described in Sect.~\ref{sec:full_curve_fit}).
It can be seen that there is no dependence of the decay- or rise-time measurements on the varying angular separations, in agreement with the results presented in Sect.~\ref{sec:decay_rise_vs_angle} and Appendix~\ref{sec:cross_correl}.
These simulations are also consistent with the outputs of \cite{2020ApJ...898...94K} who showed that the simulated time profiles did not change with angles varying between 0--50$\degr$.
In addition, even though the simulated detectors are all located at 1~au, their Euclidean distance from the source varies according to their angular separation.  Their varying positions do not appear to enforce a trend in the decay- or rise-time measurements, consistent with the results from the observations presented in Appendix~\ref{sec:decay_rise_vs_distance}.

Notably, fitting the simulated time profiles with the function in Eq.~(\ref{eqn:fit_function}) enables a direct comparison of the obtained rise and decay times.  The average predicted ratio of rise-to-decay times obtained from these simulations (at $\sim$250~kHz) is~$\sim$0.31$\pm$0.02 (at the $S_{max}/e$ level; and~$\sim$0.37$\pm$0.03 at the $S_{max}/2$ level).
It should be noted that the simulations assume a point radio source and an instantaneous photon injection into the heliosphere.  We emphasise that the point source refers only to the intrinsic size of the simulated source, and not the much larger simulated (scattered) size from which the photons observed by a detector arrive.  These two conditions mean that the value predicted by these simulations can be considered as a lower limit, since any increase in the photon-injection time or intrinsic volume may lead to a longer rise time, corresponding to a larger rise-to-decay time ratio.  Indeed, the ratio predicted from the simulations matches the lowest value calculated from observations (~$\sim$0.31
at the $S_{max}/e$ level and~$\sim$0.37 at the $S_{max}/2$ level; Sect.~\ref{sec:rise_decay_ratio}).

%=================================
\section{Frequency-dependent broadening correction} \label{sec:freq_corrections}
The frequency dependence of the decay time has the form \citep{2019ApJ...884..122K}
\begin{equation} \label{eqn:decay_vs_freq}
    \tau_{d,f} = (72.2 \pm 0.3) \, f^{-0.97 \pm 0.03} \, ,
\end{equation}
where the frequency $f$ is given in MHz.
Although the exact constants can slightly vary between studies, this particular form was chosen as \cite{2019ApJ...884..122K} combined the results of multiple studies from various instruments into a single dataset, and obtained the dependence of the decay time $\tau_{d,f}$ on frequency $f$ over a very large range of frequencies ($\sim 0.1$--$300$~MHz).
An equivalent relationship for the rise time is \citep{2018A&A...614A..69R}
\begin{equation} \label{eqn:rise_vs_freq}
    \tau_{r,f} = (1.5 \pm 0.1) \, (f/30)^{-0.77 \pm 0.14} \, ,
\end{equation}
where $f$ is given in MHz.

The relations in Eqs.~(\ref{eqn:decay_vs_freq}) and~(\ref{eqn:rise_vs_freq}) are used in this analysis to estimate the systematic offset of the decay- and rise-time measurements, respectively, stemming from the fact that the time profiles analysed for each individual event were not recorded at the exact same frequency by all spacecraft, introducing frequency-dependent broadening of the profiles.  The systematic offset $\delta \tau_\star$ for the decay and rise times (both denoted by $\tau_\star$) is estimated as
\begin{equation} \label{eqn:system_offset}
    \delta \tau_\star = \tau_{\star}(f_{av}) - \tau_{\star}(f) \, ,
\end{equation}
where $\tau_{\star}(f_{av})$ represents the expected $\tau_\star$ at the average frequency $f_{av}$ of the given event's time profiles, and $\tau_{\star}(f)$ is the expected $\tau_\star$ value for the specific frequency $f$ analysed.  These expected values are calculated using Eqs.~(\ref{eqn:decay_vs_freq}) and~(\ref{eqn:rise_vs_freq}).  The estimated systematic offset can then be added to the decay or rise time measured at a given frequency, correcting for the difference in recorded frequencies by the different spacecraft.

%=================================
\section{Decay and rise time vs distance from source} \label{sec:decay_rise_vs_distance}
We have investigated, using the Type III bursts observed by multiple spacecraft, if there exists a relationship between the distance of the spacecraft from the radio source and the measurements of decay and rise time.

\begin{figure*}[ht!]
    \centering
    \includegraphics[width=0.49\textwidth, keepaspectratio=true]{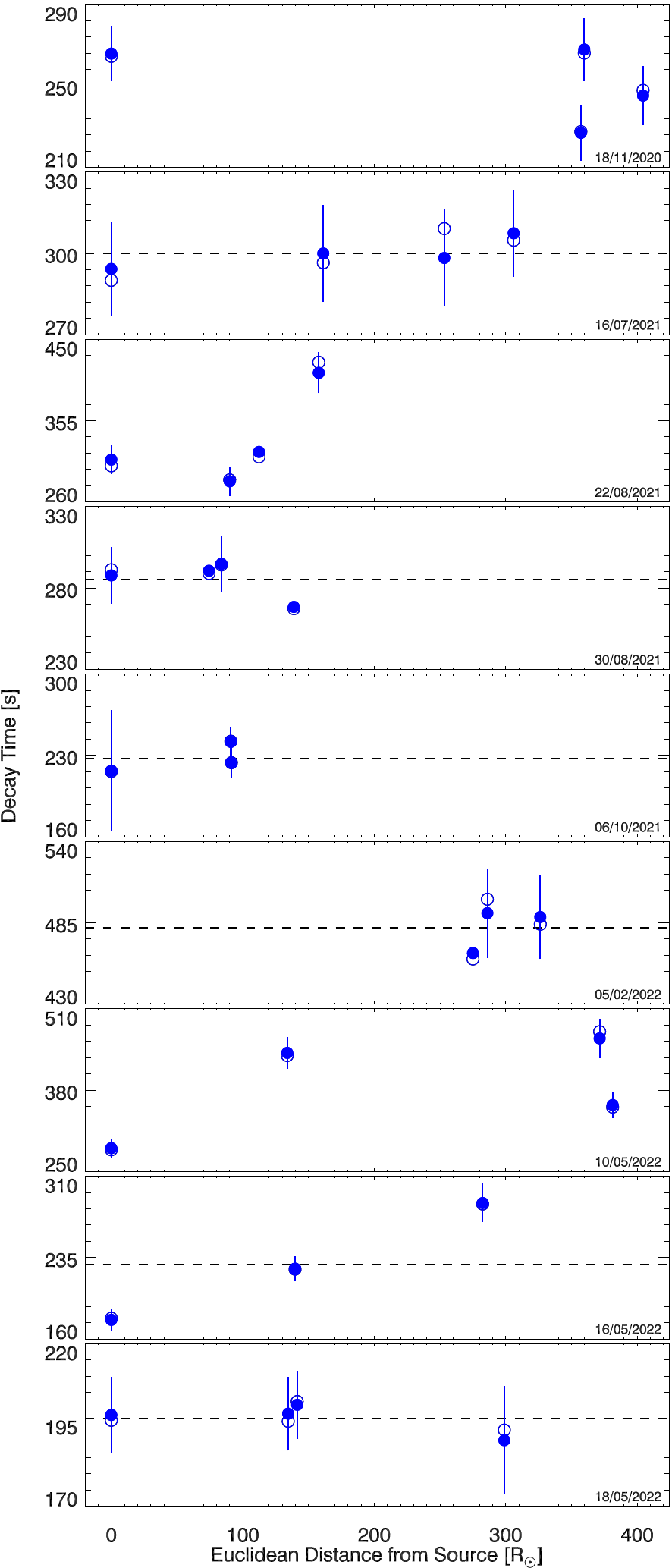}~
    \includegraphics[width=0.49\textwidth, keepaspectratio=true]{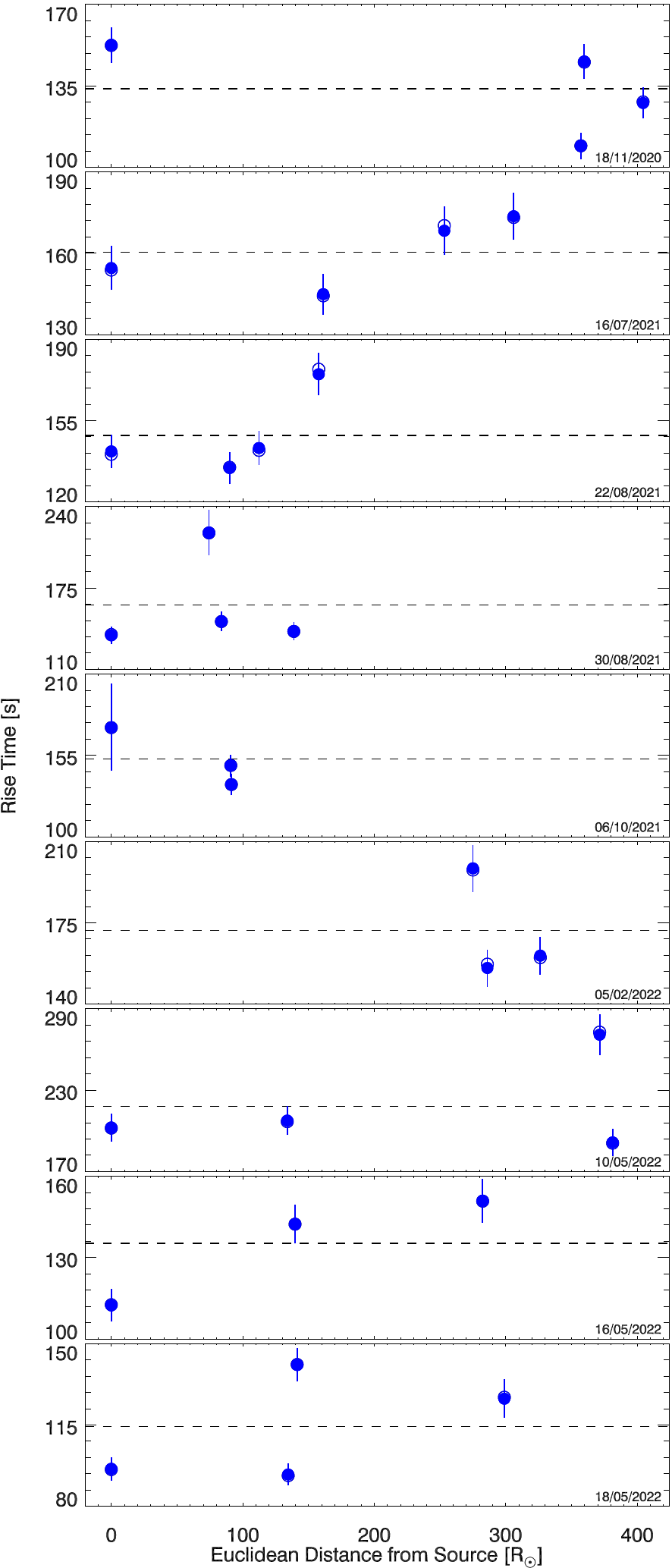}
    \caption{
    Observed decay times (left) and rise times (right), with their associated errors, as a function of the Euclidean distance of the spacecraft from the source.  Each row represents one Type III burst, and the spacecraft that recorded the Langmuir waves in situ is always located at 0~$\mathrm{R_\sun}$. The black dotted line in each panel represents the average decay/rise time value measured for the given event.  Empty circles represent the decay/rise times obtained by fitting Eq.~(\ref{eqn:fit_function}), whereas filled circles represent the same measurements but corrected for the systematic offset induced by any difference in the recorded frequencies between spacecraft (Appendix~\ref{sec:freq_corrections}).
    For clarity, error bars are only plotted for the filled circles.
    }
    \label{fig:decay_rise_vs_distance}
\end{figure*}

Figure~\ref{fig:decay_rise_vs_distance} shows the obtained decay and rise times as a function of the Euclidean distance of the spacecraft from the radio source ($R_{12}$ in Appendix~\ref{sec:angle_calculation}).
The studied events include cases where spacecraft were located on opposite sides of the Sun, separated by large angles ($\sim$180$\degr$) and Euclidean distances as large as $\sim$400~$\mathrm{R_\sun}$ ($\approx 1.85$~au).
No consistent trend or dependency between the decay- or rise-time measurements and the Euclidean distance of the detectors from the radio source is observed.  Similar to the results in Sect.~\ref{sec:decay_rise_vs_angle}, it cannot be claimed that the decay and rise times either systematically increase or decrease with increasing distance from the source.

These results contradict the suggestion by \cite{2021A&A...656A..33V} that a difference in the heliocentric distance of the spacecraft observing Type III bursts resulted in a difference (of about 45~s on average) in the decay time measurements between the two spacecraft used (WIND and SolO).  Although the study by \cite{2021A&A...656A..33V} involved a larger number of Type III events (35 in total), the measurements were not always conducted at comparable frequencies, and no correction for any resulting difference in the decay times was applied (see Sect.~\ref{sec:decay_rise_vs_angle} and Appendix~\ref{sec:freq_corrections}).  The obtained (non-corrected) decay times can be compared at similar frequencies at a time, where no conclusive difference between the spacecraft measurements can be claimed, assuming that the error bars from the decay time estimation are not ignored (see Fig.~8 in \cite{2021A&A...656A..33V}).  Additionally, the decay times of each Type III burst were only recorded by 2 spacecraft, meaning that it may be harder to distinguish an outlier compared to studies where 3 or more spacecraft observed the same event, as is the case in this work.  Moreover, \cite{2021A&A...656A..33V} estimated the decay times using a single-exponential fit, which could introduce further uncertainties and variation (Appendix~\ref{sec:func_reliability}) in the obtained decay times from one spacecraft to the other, which also depend on the quality of the data recording itself (i.e. temporal resolution and noise).
It should be noted that there exists a specific and limiting case where the decay times measured at different distances from the source can vary, even though it is not applicable in this study (see Sect.~\ref{sec:conclusion} for a discussion).

Our analysis provides a new insight into such studies, since we used more than 2 spacecraft (at least 3, and 4 in most cases; Table~\ref{tab:events_list}), and estimated the decay times using an improved fitting function (see Sect.~\ref{sec:full_curve_fit} and Appendix~\ref{sec:func_reliability}).  We have also considered the 3D locations and calculated the (physically-meaningful) Euclidean distance of the spacecraft from the radio source, instead of simply comparing their heliocentric distance which is not (necessarily) analogous to their distance from the source location.
Future multi-spacecraft studies with improved fitting methods should be conducted for a statistically larger number of events, taking the 3D position of spacecraft into account, as well as the physically-meaningful Euclidean distance of said spacecraft from the source.

%------------------------------------FIGURE-------------------------------------
%----------------------------------END FIGURE-----------------------------------

%=================================
\section{Cross-correlation of the light curves} \label{sec:cross_correl}
The results presented in Sect.~\ref{sec:decay_rise_vs_angle} are evaluated by cross-correlating the time profiles of each Type III burst recorded by the different spacecraft.
The only data processing conducted for this method is the subtraction of the (constant) background signal level (estimated as described in Appendix~\ref{sec:bg_estimation}) and the subsequent omission of the data points that fall below this background level from the dataset.

Signal measurements from each spacecraft are conducted at different times, so the light curves were aligned with respect to the peak-signal times and were (linearly) interpolated for the purposes of calculating the lag between them.
To interpolate the data, the light curve with the largest number of data points within the defined time range was taken as the base for the interpolation.  Once light curves were interpolated, the cross-correlation (and lag) between the light curves was calculated.  The closer the cross-correlation value is to 1.0, the larger the similarity is between the light curves.

Figure~\ref{fig:cross_corr} shows the (non-interpolated) light curves that were re-aligned using the obtained lag.
The depicted light curves were also normalised to the peak-signal value obtained from the fit (Eq.~\ref{eqn:fit_function}), instead of the maximum-recorded data point, in order to eliminate potential differences induced by overly-noisy data.  This leads to the peak-signals of some of the light curves in Fig.~\ref{fig:cross_corr} appearing above or below 1.

The depicted data and corresponding cross-correlation values indicate that the shape of the radio burst light curves (and thus the decay and rise time) does not vary significantly at comparable frequencies with varying observer positions, nor with varying distances of the spacecraft from the source.
Despite the noise in the data, the similarity between the light curves is high, often exceeding 95\%.  This is the case even though lower signals (near the background-noise level), which tend to be noisier, were considered in the calculation.
Therefore, these results corroborate the outcomes of the fitting function proposed in Sect.~\ref{sec:full_curve_fit}, seen in Fig.~\ref{fig:decay_rise_vs_angle}.

%=================================
\section{Decay and rise time vs frequency} \label{sec:decay_rise_vs_freq}
\begin{figure}[ht!]
    \centering
    \includegraphics[width=0.5\textwidth, keepaspectratio=true]{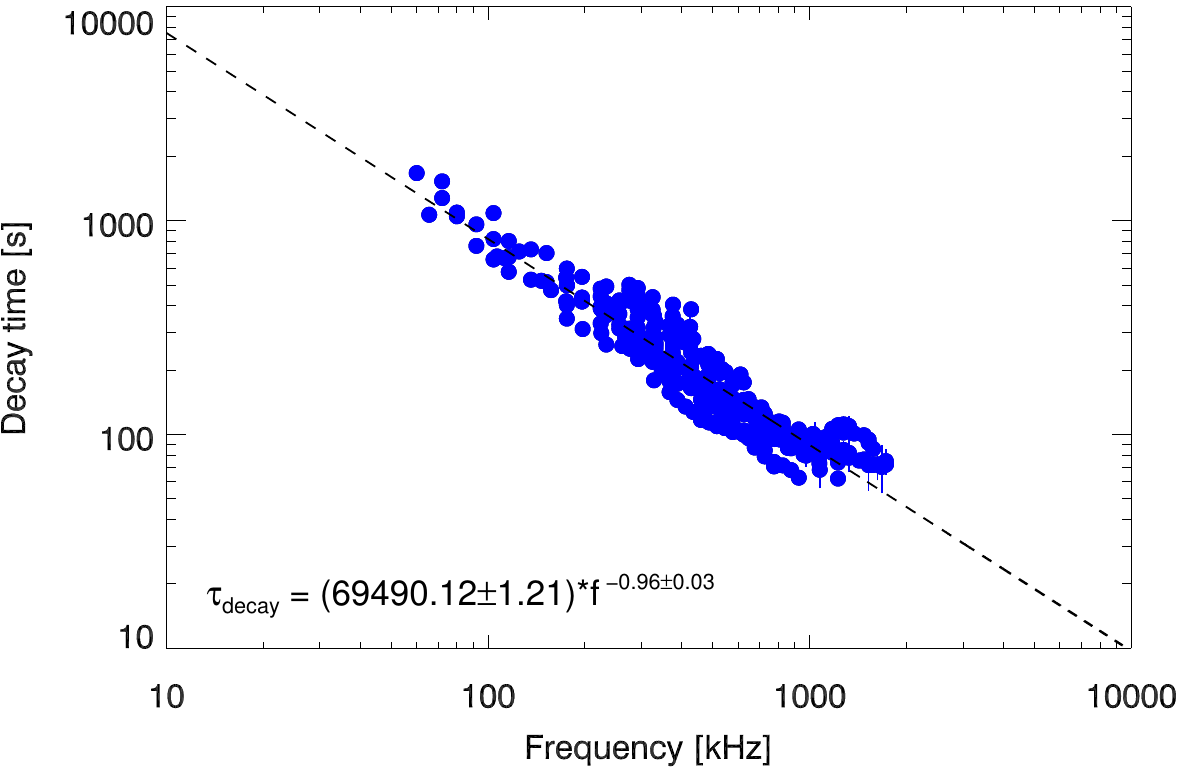}
    \includegraphics[width=0.5\textwidth, keepaspectratio=true]{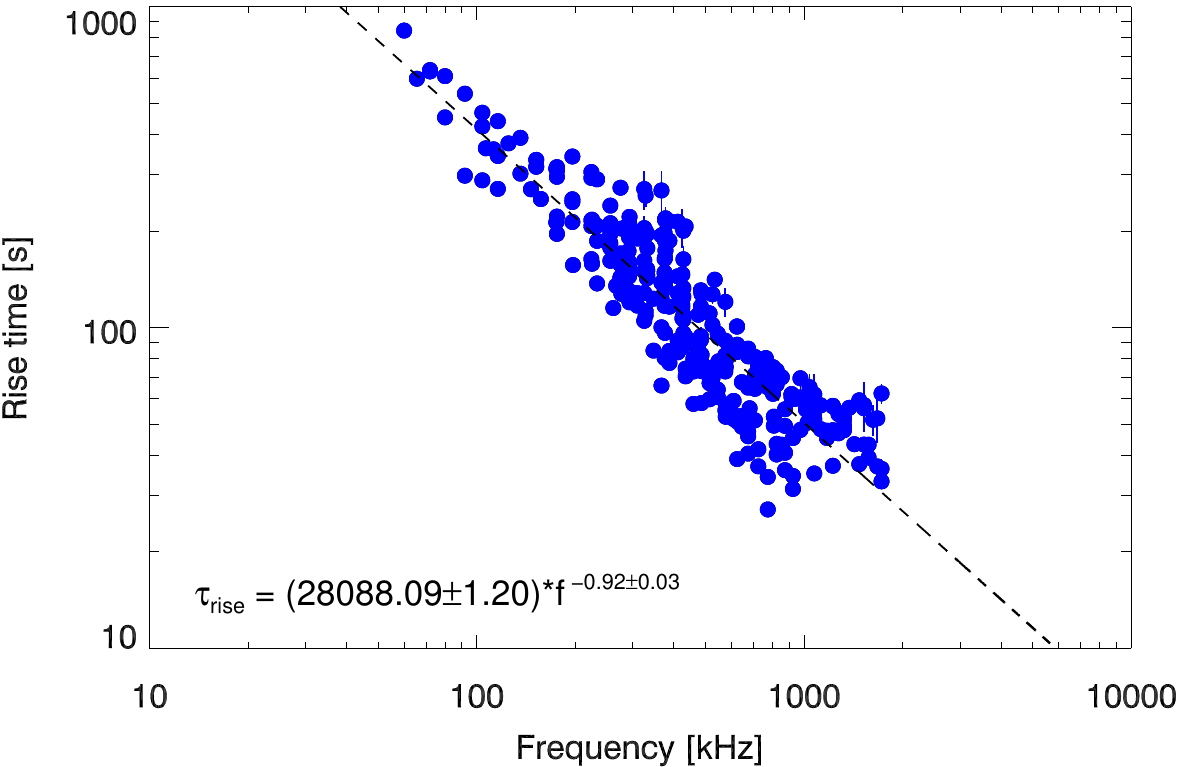}
    \caption{Decay times (top) and rise times (bottom) as a function of frequency, for 307 measurements at frequencies ranging from 60--1725~kHz, obtained from the Type III bursts listed in Table~\ref{tab:events_list}.  The values and their uncertainties were calculated at the $S_{max}/e$ level as described in Sect.~\ref{sec:full_curve_fit}.  The black dashed lines represent the weighted linear fit to the data, whose exact form is shown on the bottom left of each panel.  It should be noted that the frequencies shown here (and thus reflected in the obtained relations) are given in kHz.
    }
    \label{fig:decay_and_rise_vs_freq}
\end{figure}

Figure~\ref{fig:decay_and_rise_vs_freq} shows 307 measurements of decay and rise times obtained from the selected Type III bursts (and calculated using Eq.~(\ref{eqn:fit_function})) against the frequency $f$ they were observed, for frequencies ranging from 60--1725~kHz.
The obtained relationship between the estimated decay times and frequency is
\begin{equation} \label{eqn:my_decay_vs_freq}
    \tau_{d} \propto f^{-0.96 \pm 0.03} \, ,
\end{equation}
consistent with those stated in the literature (approximately $\propto 1/f$), including Eq.~(\ref{eqn:decay_vs_freq}).
Equivalently, the relation between the rise times and frequency is found to be
\begin{equation} \label{eqn:my_rise_vs_freq}
    \tau_{r} \propto f^{-0.92 \pm 0.03} \, ,
\end{equation}
also consistent with previous results, like Eq.~(\ref{eqn:rise_vs_freq}).
The successful reproduction of the decay- and rise-time relationship to frequency corroborates the validity of Eq.~(\ref{eqn:fit_function}).

Generally, rise times have not been sufficiently studied.  By extension, the frequency-dependence of rise times has not been studied to the extend of that of decay times.  This is especially true at lower frequencies where the available temporal resolution of spacecraft may have prohibited the ability to confidently fit the rise phase of time profiles with a single exponential function.  This is another advantage of the proposed fitting function in Eq.~(\ref{eqn:fit_function}), which can be less sensitive to a lower number of rise-phase measurements, given that it is dependent on the measurements constituting the entire time profile, including the longer (and thus better-resolved) decay phase.

%------------------------------------FIGURE-------------------------------------
%----------------------------------END FIGURE-----------------------------------

\end{appendix}

\end{document}